\documentclass[prl,reprint,twocolumn,showpacs,superscriptaddress]{revtex4-1}
\usepackage{bm,mathrsfs}
\usepackage[dvipdfmx]{graphicx}
\usepackage{epsfig}
\usepackage{amsmath,bbm}
\usepackage{amsfonts,amssymb,physics}
\usepackage{times}
\usepackage{dsfont}
\usepackage{enumitem}  
\usepackage{comment}
\usepackage[dvipdfmx,colorlinks=true,linkcolor=blue,urlcolor=blue,citecolor=blue]{hyperref}
\newcommand{\beq}{\begin{equation}}
\newcommand{\eeq}{\end{equation}}
\newcommand{\bseq}{\begin{subequations}}
\newcommand{\eseq}{\end{subequations}}
\newcommand{\bal}{\begin{aligned}[b]}
\newcommand{\eal}{\end{aligned}}
\newcommand{\beqa}{\begin{eqnarray}}
\newcommand{\eeqa}{\end{eqnarray}}

\begin{document}

\title{Berezinskii-Kosterlitz-Thouless phase transition 
with Rabi-coupled bosons}

\author{Koichiro Furutani}
\email{koichiro.furutani@phd.unipd.it}
\affiliation{Dipartimento di Fisica e Astronomia 'Galileo Galilei'  
and QTech Center, Universit\`a di Padova, via Marzolo 8, 35131 Padova, Italy}
\affiliation{Istituto Nazionale di Fisica Nucleare, Sezione di Padova, 
via Marzolo 8, 35131 Padova, Italy}
\author{Andrea Perali}
\affiliation{School of Pharmacy, Physics Unit, 
Universit\`a di Camerino, Via Madonna delle Carceri 9, 62032 Camerino, Italy}
\author{Luca Salasnich}
\affiliation{Dipartimento di Fisica e Astronomia 'Galileo Galilei'  
and QTech Center, Universit\`a di Padova, via Marzolo 8, 35131 Padova, Italy}
\affiliation{Istituto Nazionale di Fisica Nucleare, Sezione di Padova, 
via Marzolo 8, 35131 Padova, Italy}
\affiliation{Istituto Nazionale di Ottica del Consiglio Nazionale delle Ricerche, 
via Carrara 2, 50019 Sesto Fiorentino, Italy}

\begin{abstract}
We theoretically investigate the superfluid-normal-state Berezinskii-Kosterlitz-Thouless transition in a binary mixture of bosonic atoms with Rabi coupling under balanced densities. 
We find the nonmonotonic behavior of the transition temperature with respect to the intercomponent coupling and amplification of the transition temperature for finite values of Rabi coupling, but for small intracomponent couplings. 
We develop the Nelson-Kosterlitz renormalization-group equations in the two-component Bose mixture and obtain the Nelson-Kosterlitz criterion modified by a fractional parameter, which is responsible for half-integer vortices, and by Rabi coupling. 
Adopting the renormalization-group approach, we clarify the dependence of the Berezinskii-Kosterlitz-Thouless transition temperature on the Rabi coupling and the intercomponent coupling. 
Analysis of the first and second sound velocities also reveals the suppression of quasicrossing of the two sound modes with a finite Rabi coupling in the low-temperature regime. 
Our results for a two-dimensional binary Bose superfluid contribute to the understanding of a broad range of multicomponent quantum systems such as two-dimensional multiband superconductors. 
\end{abstract}
\date{\today}

\maketitle

The Berezinskii-Kosterlitz-Thouless (BKT) transition is one of the most striking phenomena that occur in a two-dimensional (2D) superfluid realized in thin films of $^{4}\mathrm{He}$ \cite{kadanoff,clow,reppy,amit,bergman,henkel,chester72,chester73,chan,berthold,bishop78,bishop80,bishop81,kotsubo,minnhagen,agnolet}, ultracold atoms in a planar geometry \cite{hadzibabic2006,hadzibabic2015,phillips,cornell,turlapov,dyke,feld,zwierlein,kohl,dalibard2011,dalibard2012,chin2011,chin2013,vogt,bohlen,ries,murthy,ville,lompe,hadzibabic2021,pitaevskii,svistunov} or in a spherical bubble trap \cite{tononi19,tononi20,tononi22,lundblad}, and exciton-polariton systems \cite{boyce,willander,brameti,bloch,ciuti,yamamoto,caputo,szymanska}.
The BKT transition originates from unbindings of vortex-antivortex pairs, and a proliferation of free vortices and antivortices \cite{berezinskii,kosterlitz,nelson}. 
It was first experimentally observed in thin $^{4}\mathrm{He}$ films \cite{bishop78} and later also in superconducting films \cite{kadin,fiory,xue,zhao,sharma2022}, ultracold atomic gases \cite{hadzibabic2006,hadzibabic2015,hadzibabic2021,phillips,cornell,chin2011,chin2013,dalibard2011,dalibard2012,ries,murthy,ville}, and exciton-polariton systems \cite{yamamoto,caputo}. 
A BKT transition to electron-hole superfluidity in 2D atomic double layers has been also predicted and is under current investigation \cite{perali2013,wang2019}. 
A stark contrast to three-dimensional (3D) superfluidity is a discontinuous jump of the superfluid density at the BKT transition temperature in a 2D superfluid \cite{nelson,mora, ozawa,salasnich2016,miki,furutani,singh,singh22}. 
It also leads to a jump of the second sound velocity, which was experimentally measured recently with a $^{39}\mathrm{K}$ atomic gas \cite{hadzibabic2021}. 
To theoretically investigate the BKT transition, there are mainly two approaches. 
One is universal relations which are valid in the vicinity of the BKT transition temperature \cite{prokofev2001,prokofev2002,dalibard2011,chin2011,dupuis}. 
The other approach is to use the Nelson-Kosterlitz (NK) renormalization group (RG) equations, which are responsible for RG flows of the vortex fugacity and the phase stiffness associated with the superfluid density \cite{nelson}. 
An advantage of the RG approach is that it is also valid in the low-temperature regime. 

In contrast to a single-component Bose gas, a multicomponent Bose mixture has significant qualitative differences such as the Andreev-Bashkin entrainment effect between different species \cite{andreev,svistunov,fil,recati2017,konietin,enss,shin2020,recati2021}, the emergence of fractional circulation of vorticity \cite{son2002,mueller,kasamatsu2003,kasamatsu2004,kasamatsu2005,kasamatsu2009,wei,kuo,eto2011,eto2012,eto2013,nitta2013L,nitta2013A,dantas,kasamatsu2016,stringari2016,eto2018,lamacraft}, and the modification of the NK criterion \cite{cross1979,korshunov1984}. 
There are also several theoretical analyses of the BKT transition in a bilayer $XY$ model \cite{granato1986,bighin2019}, which has similarities to 2D binary Bose mixtures, and a Monte Carlo simulation in a binary Bose mixture with finite Rabi coupling \cite{nitta}. 
Finite Rabi coupling makes half-quantized vortices, which are vortices in one of the two components of the Bose atoms, topologically unstable but makes vortex molecules, which consist of two vortices of both components with positive or negative charges, stable. 
Reference \cite{nitta} proposed that the topological excitations that induce the BKT transition are also replaced with vortex molecule-antimolecule pairs instead of vortex-antivortex pairs. 
Renormalization group analysis taking into account these distinct topological excitations is crucial to predict physical quantities such as sound velocities and provide a coherent understanding of multicomponent superfluidity. 

In this Letter, we consider a 2D atomic Bose gas confined in a quadratic region of area $L^2$, at temperature $T$, and with a chemical potential $\mu$ across the BKT transition temperature through the RG approach. 
The bosonic gas is characterized by atoms with two hyperfine components in their energy-level spectrum. 
In addition to the usual intraspecies ($g=g_{11}=g_{22}>0$) and interspecies ($g_{12}$) contact interactions, atoms in different hyperfine states interact via an external coherent Rabi coupling of frequency $\omega_{\rm R}(\ge0)$, which drives an exchange of atoms between the two components. 
The presence of the Rabi coupling implies that only the total number $N=N_1+N_2$ of atoms is conserved, with $N_{a=1,2}$ being the number of atoms in the $a$th hyperfine component. 
The existence and stability of the ground state with balanced densities $N_1=N_2$ were extensively discussed in Refs.~\cite{abad2013,bertacco2017}. 
We focus on the balanced and uniform ground state throughout this Letter. 

Our two-component Bose-atom systems are a counterpart to strongly coupled multiband superconductors in which all the partial condensates are close to the Bose-Einstein condensation regime. 
The Rabi coupling corresponds in multiband superconductors to the Cooper-pair exchange among different bands and even in the case of multiband systems, it is the total number of carriers that is conserved, with redistribution of densities among the bands depending on the parameter configuration and on the renormalization of the chemical potential \cite{perali2014,perali2019,pieri}. 
Hence, the present investigation of Rabi coupled bosons can shed light on the BKT transition and collective modes in 2D multiband superconductors, a growing field of study for their fundamental interest and quantum technology applications \cite{perali2015}.

We first examine the two branches of elementary excitations, which are related to Rabi coupling and intercomponent coupling. 
To consider the BKT transition, we develop NK RG equations in the two-component Bose gas. 
We point out that the NK criterion that provides the BKT transition temperature is modified due to the fractional parameter. 
The fractional parameter is also responsible for the half circulation of vorticity in a population-balanced binary Bose mixture. 
With finite Rabi coupling, on the other hand, the NK criterion reduces to the one in the single-component case related to the formation of vortex molecule-antimolecule pairs.  
This modification of the NK criterion is also consistent with previous theoretical predictions based on Monte Carlo analysis under balanced densities \cite{nitta}. 
We investigate the dependence of the BKT transition temperature on Rabi coupling and inter-component coupling. 
It shows a nonmonotonic behavior with respect to the intercomponent coupling and amplifies the maximum transition temperature for each value of Rabi coupling. 
Finally, we determine the first and second sound velocities across the BKT transition temperature. 
We confirm the jump of the second sound velocity at the BKT transition temperature. 
At low temperatures, in particular, finite Rabi coupling is found to hinder quasicrossing behavior due to the presence of a gapped mode, in contrast to the single-component superfluids \cite{LeeYang,griffin,taylor,hu,stringari,furutani}.

The Bogoliubov spectrum of elementary excitations in a uniform system has two branches given by \cite{abad2013,bertacco2017}
\beqa
E^{(-)}_{k} &=& 
\sqrt{\varepsilon_{k}\left[\varepsilon_{k}+2\left(\mu +\hbar\omega_{\rm R}\right)\right]}, 
\label{b1}
\\
E^{(+)}_{k} &=&
\sqrt{\varepsilon_{k}\left(\varepsilon_{k}+2A\right)+B},
\label{b2}
\eeqa
with $\varepsilon_{k}=\hbar^{2}k^{2}/(2m)$ and $m$ being the atomic mass. 
We set $\eta=g_{12}/g$, and the two parameters appearing in Eq.~\eqref{b2} are 
\beqa
A  &=& \frac{1-\eta}{1+\eta}\left(\mu+\hbar\omega_{\rm R}\right) + 
2\hbar\omega_{\rm R} ,
\\
B &=& 4\hbar\omega_{\rm R} \left[\frac{1-\eta}{1+\eta}
\left(\mu + \hbar\omega_{\rm R}\right) 
+ \hbar\omega_{\rm R}\right] \; .
\eeqa
At the mean-field level, for the uniform ground state with balanced densities, the chemical potential $\mu$ reads \cite{abad2013,bertacco2017} 
\beq 
\mu = \frac{1+\eta}{2}gn -\hbar\omega_{\rm R} \; , 
\label{echem-mf}
\eeq
where $n=N/L^2$ is the 2D total number density of bosons. 
The uniform ground state with balanced densities, characterized by $n_1=n_2=n/2$, is stable under the conditions $g + g_{12} > 0$ and $(g-g_{12})n+2\hbar\omega_{\rm R} > 0$ 
\cite{abad2013,bertacco2017}, namely, $-1< \eta < 1 +2\hbar\omega_{\rm R}/(gn)$ 
with $g>0$. 
By using Eq.~\eqref{echem-mf}, parameters $A$ and $B$ become $A=gn(1-\eta)/2+2\hbar \omega_{\rm R}$ and $B=4\hbar\omega_{\rm R}\left[gn(1-\eta)/2+\hbar\omega_{\rm R}\right]$. 
For small wavenumbers, the elementary excitations in Eqs.~\eqref{b1} and \eqref{b2} read $E^{(-)}_{k}=c_{\rm B}\hbar k$ and $E^{(+)}_{k}=\sqrt{B}+\varepsilon_{k}A/\sqrt{B}$, showing explicitly that the mode $E^{(-)}_{k}$ is gapless while the mode $E^{(+)}_{k}$ is gapped (if $\omega_{\rm R}\neq 0$). 
Notice that $c_{\rm B}=[gn(1+\eta)/(2m)]^{1/2}$ is the Bogoliubov speed of sound for the uniform system. 
For $\eta=1$, one recovers the familiar expression $c_{\rm B}=\sqrt{gn/m}$.

By adopting Landau's approach \cite{landau}, at finite temperature $T$, the superfluid density of the system is given by 
\beq 
n_{\rm s}^{(0)}(T) = n - n_{\rm n}^{(-)}(T) - n_{\rm n}^{(+)}(T) ,
\label{ns0}
\eeq
where 
\beq
n_{\rm n}^{(\pm)}(T) = - {1\over 2} \int 
{d^2{\bm k}\over (2\pi)^2} {\hbar^2k^2\over 2m} f_{T}'(E^{(\pm)}_{k}) 
\label{nnplusminus}
\eeq
is the thermally activated normal density due to the elementary excitations. 
In the formula, $f_{T}'(E)$ is the derivative with respect to $E$ of the Bose distribution function $f_{T}(E)=1/[e^{E/(k_{\rm B}T)}-1]$, with $k_{\rm B}$ being the Boltzmann constant.

It is important to stress that the superfluid density obtained in Eq.~\eqref{ns0} does not take into account the formation of quantized vortices. 
The bare superfluid density $n_{\rm s}^{(0)}(T)$ goes to zero at a critical temperature that is larger than $T_{\rm c}$, the critical temperature of the BKT phase transition induced by the unbinding of vortex-antivortex pairs and the proliferation of free quantized vortices described by NK RG equations \cite{berezinskii,kosterlitz}. 
In a single-component 2D Bose gas, the NK RG equations are given by \cite{nelson,giamarchi,altlandsimons,supplement}
\beq
\partial_{l}K(l)^{-1}=4\pi^{3}y(l)^{2}, \quad
\partial_{l}y(l)=\left[2-\pi K(l)\right]y(l),
\label{RGeqs}
\eeq
with $K(l)\equiv\hbar^{2}n_{\mathrm{s}}^{(l)}(T)/(mk_{\mathrm{B}}T)=J(l)/(k_{\rm B}T)$, $J(l)=\hbar^{2}n_{\rm s}^{(l)}(T)/m$ being the phase stiffness, and $y(l)\equiv \mathrm{exp}[-\mu_{\mathrm{v}}(l)/(k_{\mathrm{B}}T)]$, where $\mu_{\mathrm{v}}(l)$ is the vortex chemical potential at the dimensionless scale $l$. 
The BKT critical temperature $T_{\rm c}^{(0)}$ can be obtained by using 
the NK criterion which provides a fixed point of Eqs.~\eqref{RGeqs} \cite{nelson}. 
According to this criterion, $T_{\rm c}^{(0)}$ is given by the implicit formula 
\beq 
k_{\rm B}T_{\rm c}^{(0)} = {\pi\hbar^2\over 2m} n_{\rm s}(T_{\rm c}^{(0)}) \; . 
\label{criterion}
\eeq

In a binary Bose mixture with balanced densities $\alpha_{a=1,2}=n_{a}/n=1/2$; in contrast, we can obtain the following set of NK RG equations \cite{nitta,giamarchi,altlandsimons,supplement}
\bseq
\beq
\partial_{l}K(l)^{-1}=4\pi^{3}\Theta(\omega_{\rm R}) y(l)^{2},
\eeq
\beq
\partial_{l}y(l)=\left[2-\pi\Theta(\omega_{\rm R})K(l)\right]y(l), 
\eeq
\label{halfRGeqs}
\eseq
where $\Theta(x)$ is the Heaviside step function with $\Theta(0)=1/2$. 
It can be derived from the microscopic Lagrangian as in the single-component case. 
For the details of the derivation, see the Supplemental Material \cite{supplement}. 
The RG equations \eqref{halfRGeqs} give the modified NK criterion
\beq
k_{\rm B}T_{\rm c}=\frac{\pi\hbar^{2}}{2m}\Theta(\omega_{\rm R})n_{\rm s}(T_{\rm c}) 
\label{halfcriterion}
\eeq
at the BKT critical temperature $T_{\rm c}$. 
This NK criterion \eqref{halfcriterion} is consistent with the Monte Carlo analysis in Ref.~\cite{nitta}. 
To calculate the RG flow, we use the initial conditions $K(0)=\hbar^{2}n_{\mathrm{s}}^{(0)}(T)/(mk_{\mathrm{B}}T)$ and $\mu_{\mathrm{v}}(0)=\pi^{2}\Theta(\omega_{\rm R})J(0)/4$ \cite{bighin,nylen,stoof,duan}, where $n_{\mathrm{s}}^{(0)}(T)$ is calculated using Eq.~\eqref{ns0} with Eqs.~\eqref{b1}, \eqref{b2}, and \eqref{nnplusminus}. 
The maximum value of the RG scale is related to the system size as $l_{\text{max}}=\ln{(L/\xi)}$, with $\xi=\hbar/\sqrt{2mg(n/2)}$ being the vortex core size. 
Here, we note that the higher-order derivative terms in the $XY$ model can lead to corrections in the initial conditions for the RG flow. 
Indeed, it has been pointed out that the higher-order corrections are important for quantitatively accurate predictions of the BKT transition in $XY$ models in particular for a small vortex chemical potential \cite{maccari2020}. 
In our model of a binary Bose mixture, such a higher-order term of the superfluid velocity can arise and determine a quantitative change in our results with a small vortex chemical potential as well. 
In this Letter, however, since they are expected to produce moderate quantitative changes, we do not consider the effects of the spin-wave excitations on the vortex excitations, which will be the subject of a future investigation including the functional RG analysis \cite{maccari2020,kobayashi2020}.

The modification of the NK criterion in the absence of Rabi coupling reflects the half circulation of vorticity.  
Indeed, the circulation of vorticity is given by \cite{nitta}
\beq
\bal
\kappa
&\equiv \oint d\bm{s}\cdot\bm{v}_{\rm s}
=\frac{\hbar}{m}\oint d\bm{s}\cdot\frac{\abs{\psi_{1}}^{2}\grad\theta_{1}+\abs{\psi_{2}}^{2}\grad\theta_{2}}{\abs{\psi_{1}}^{2}+\abs{\psi_{2}}^{2}} ,
\eal
\eeq
with $\psi_{a=1,2}$ being the $a$th complex bosonic field, where $\bm{v}_{\rm s}$ is the superfluid velocity associated with the superfluid phase $\theta_{a=1,2}$, and $\bm{s}$ is the vector along the closed path enclosing vortices. 
With fractional parameters $\alpha_{a}=n_{a}/n$, for instance, each of the circulations for vortices $(\psi_{1},\psi_{2})\sim(\sqrt{n_{1}}e^{\pm i\theta_{0}},\sqrt{n_{2}})$, with $\theta_{0}=\arctan{(y/x)}$, is given by $\kappa_{1}=\pm 2\pi\alpha_{1}\hbar/m $ \cite{nitta}. 
For a population-balanced system $n_{1}=n_{2}=n/2$; in particular, $\alpha_{1,2}=1/2$ gives rise to half vortices. 
In the presence of Rabi coupling, on the other hand, topological defects that lead to a BKT transition are replaced with vortex molecule-antimolecule pairs instead of vortex-antivortex pairs \cite{nitta,kasamatsu2004,son2002}. 
The formation of vortex molecule pairs modifies the RG equations as in Eqs.~\eqref{halfRGeqs}, which recover the ones for the single-component case in Eqs.~\eqref{RGeqs}. 

\begin{figure}[t]
\begin{minipage}[b]{0.49\linewidth}
\centering
\includegraphics[keepaspectratio,scale=0.19]{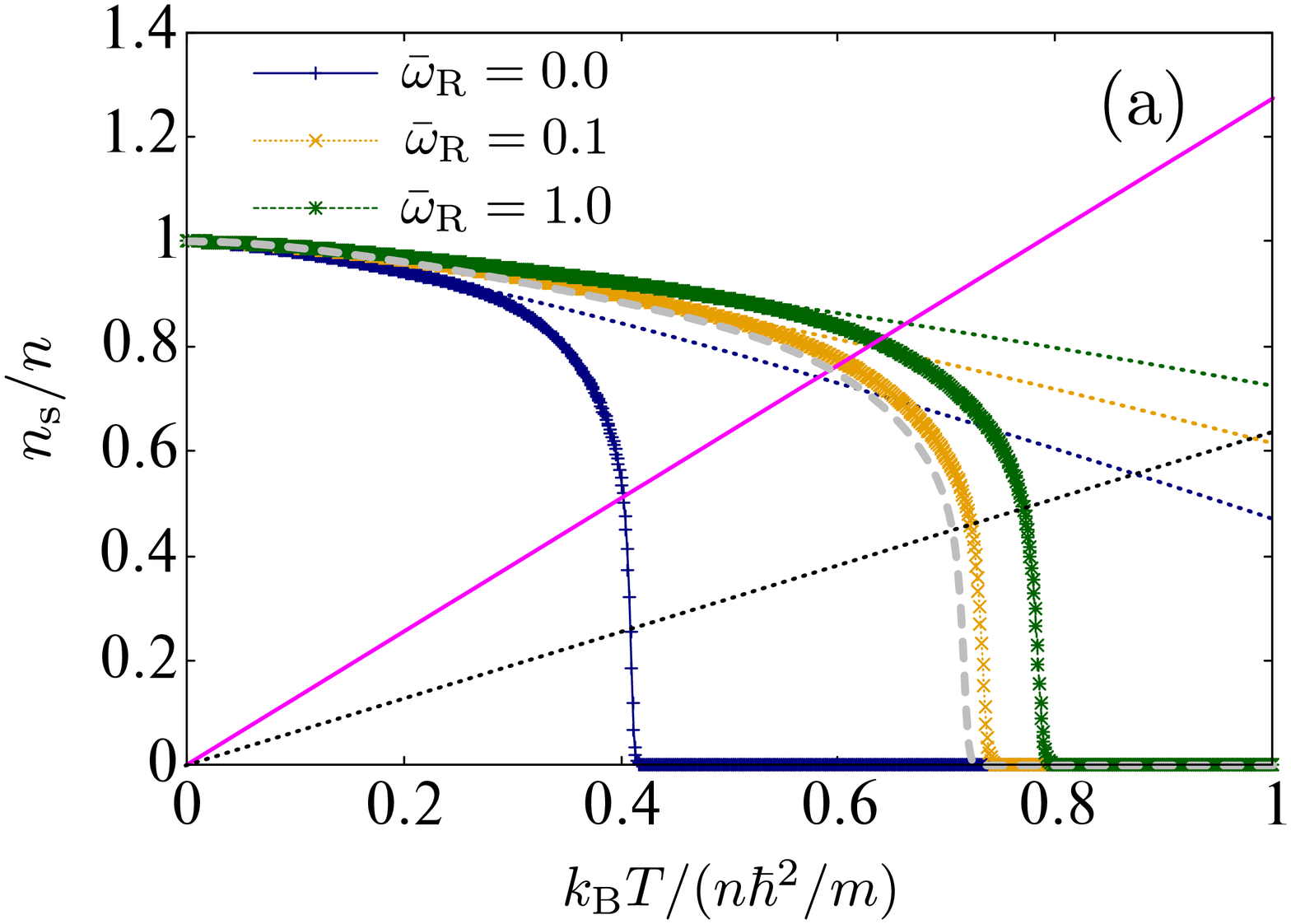}
\end{minipage}
\begin{minipage}[b]{0.49\linewidth}
\centering
\includegraphics[keepaspectratio,scale=0.20]{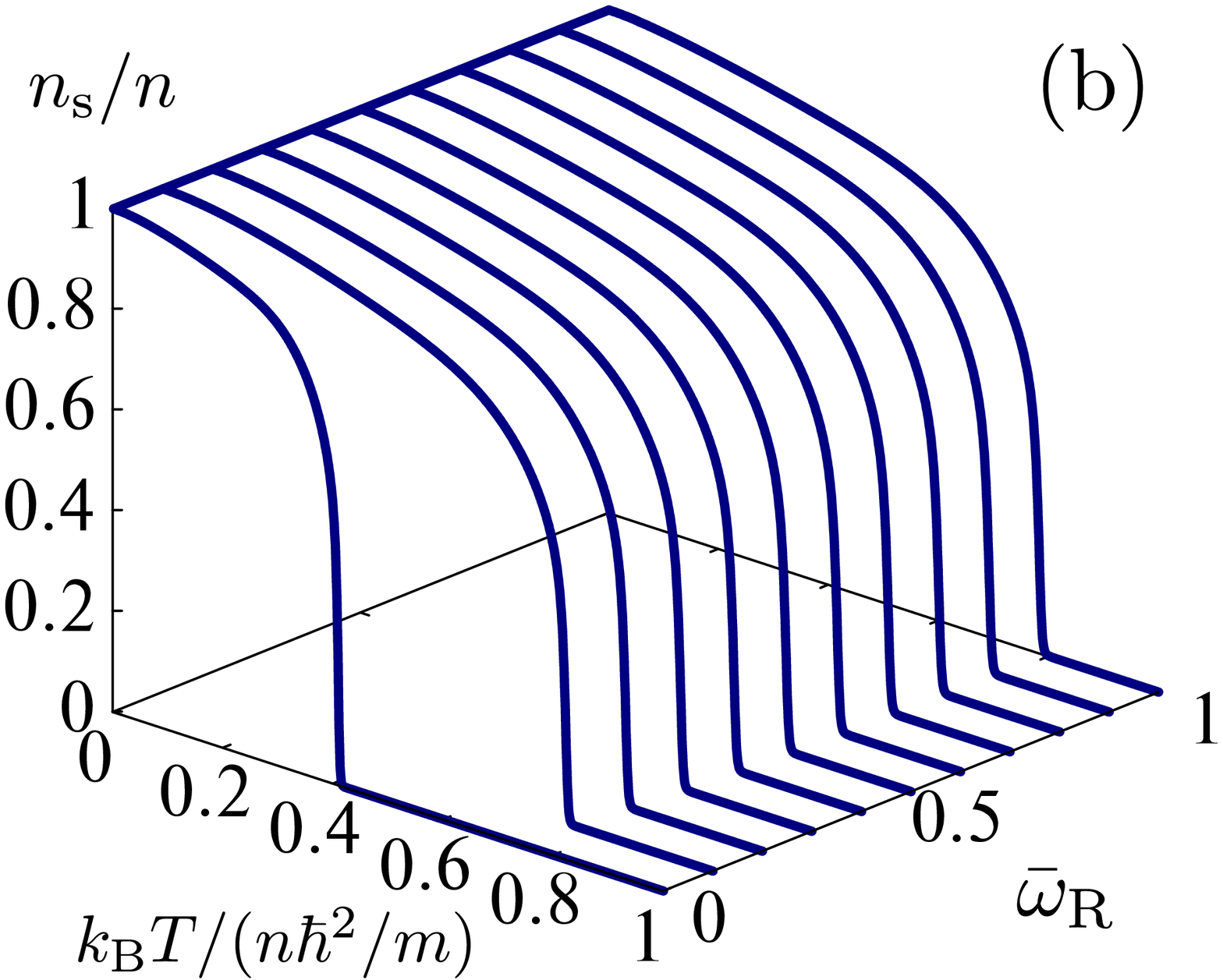}
\end{minipage}
\caption{Renormalized superfluid fraction calculated with Eqs.~\eqref{halfRGeqs} for $\Tilde{g}=mg/\hbar^{2}=0.1$ and $\eta=0$. 
(a) displays the results with $L/\xi=200$ and $\bar{\omega}_{\rm R}=\hbar\omega_{\rm R}/(n\hbar^{2}/m)=0.0,0.1,1.0$. 
The horizontal axis is the dimensionless temperature $2\pi/(n\lambda_{T}^{2})=k_{\rm B}T/(n\hbar^{2}/m)$. 
The gray dashed curve stands for the superfluid fraction in a single-component Bose gas with $\Tilde{g}=0.1$ calculated with Eqs.~\eqref{RGeqs}. 
The thin dotted curves represent the bare superfluid fraction given by Eq.~\eqref{ns0}. 
The thin solid line and thin dotted line stand for $k_{\rm B}T=\pi\hbar^{2}n_{\rm s}(T)/(4m)$ and $k_{\rm B}T=\pi\hbar^{2}n_{\rm s}(T)/(2m)$, respectively. 
(b) shows the 3D plot of the superfluid fraction as a function of the temperature and Rabi coupling.}
\label{Fignsrg0e1}
\end{figure}

Figure \ref{Fignsrg0e1} shows the renormalized superfluid fraction computed with Eqs.~\eqref{halfRGeqs} for $\Tilde{g}=mg/\hbar^{2}=0.1$ and $\eta=0$ with $L/\xi=200$. 
Figure \ref{Fignsrg0e1}(a) displays the results with $\bar{\omega}_{\mathrm{R}}=\hbar\omega_{\rm R}/(n\hbar^{2}/m)=0,0.1,1.0$. 
The horizontal axis is the dimensionless temperature $k_{\rm B}T/(n\hbar^{2}/m)=2\pi/(n\lambda_{T}^{2})$, with $\lambda_{T}=[2\pi\hbar^{2}/(mk_{\rm B}T)]^{1/2}$ being the thermal wavelength. 
The thin dotted curves stand for the bare superfluid fraction given by Eq.~\eqref{ns0}. 
Due to the finite size, the discontinuity of the renormalized superfluid fraction in the thermodynamic limit $L\to\infty$ is smeared and altered to a continuous drop \cite{supplement}. 
In the single-component case plotted by the dashed curve, the superfluid fraction intersects with the thin dotted line for $k_{\mathrm{B}}T=\pi\hbar^{2}n_{\mathrm{s}}/(2m)$ at the BKT transition temperature as in Eq.~\eqref{criterion} in the thermodynamic limit. 
In contrast, in a population-balanced binary Bose mixture, the superfluid fraction should intersect with the thin solid line for $k_{\mathrm{B}}T=\pi\hbar^{2}n_{\mathrm{s}}/(4m)$ in the absence of Rabi coupling at the BKT transition temperature as in Eq.~\eqref{halfcriterion} in the thermodynamic limit. 
With finite Rabi coupling, on the other hand, the superfluid fraction intersects with the thin dotted line for $k_{\mathrm{B}}T=\pi\hbar^{2}n_{\mathrm{s}}/(2m)$ at the BKT transition temperature in the thermodynamic limit as in the single-component Bose gas. 
A larger value of Rabi coupling shifts the transition temperature to a higher one. 
Figure \ref{Fignsrg0e1}(b) shows a 3D plot of the renormalized superfluid fraction as a function of the Rabi coupling and the temperature. 

\begin{figure}[t]
\centering
\includegraphics[keepaspectratio,scale=0.3]{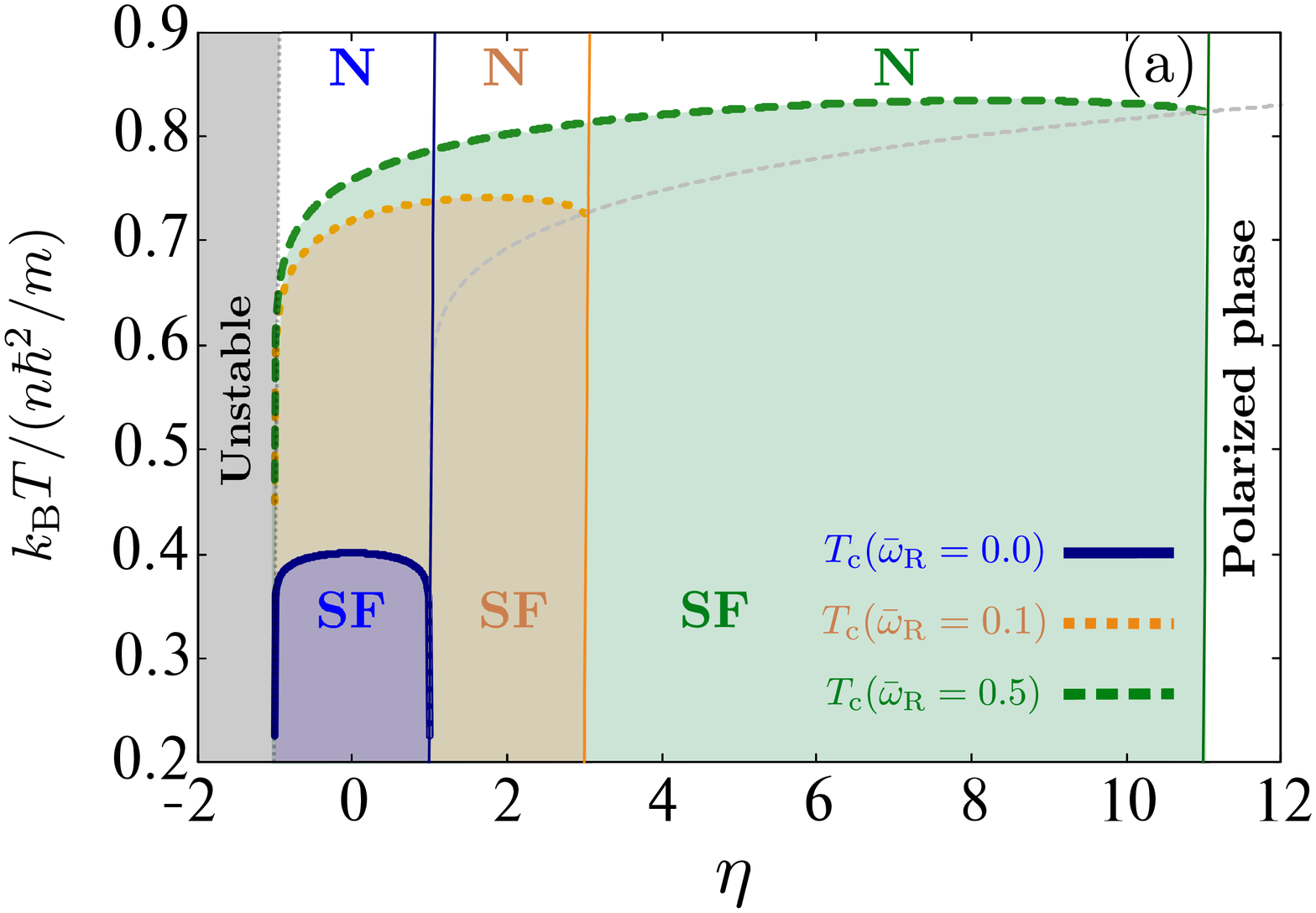}
\begin{minipage}[b]{0.49\linewidth}
\centering
\includegraphics[keepaspectratio,scale=0.20]{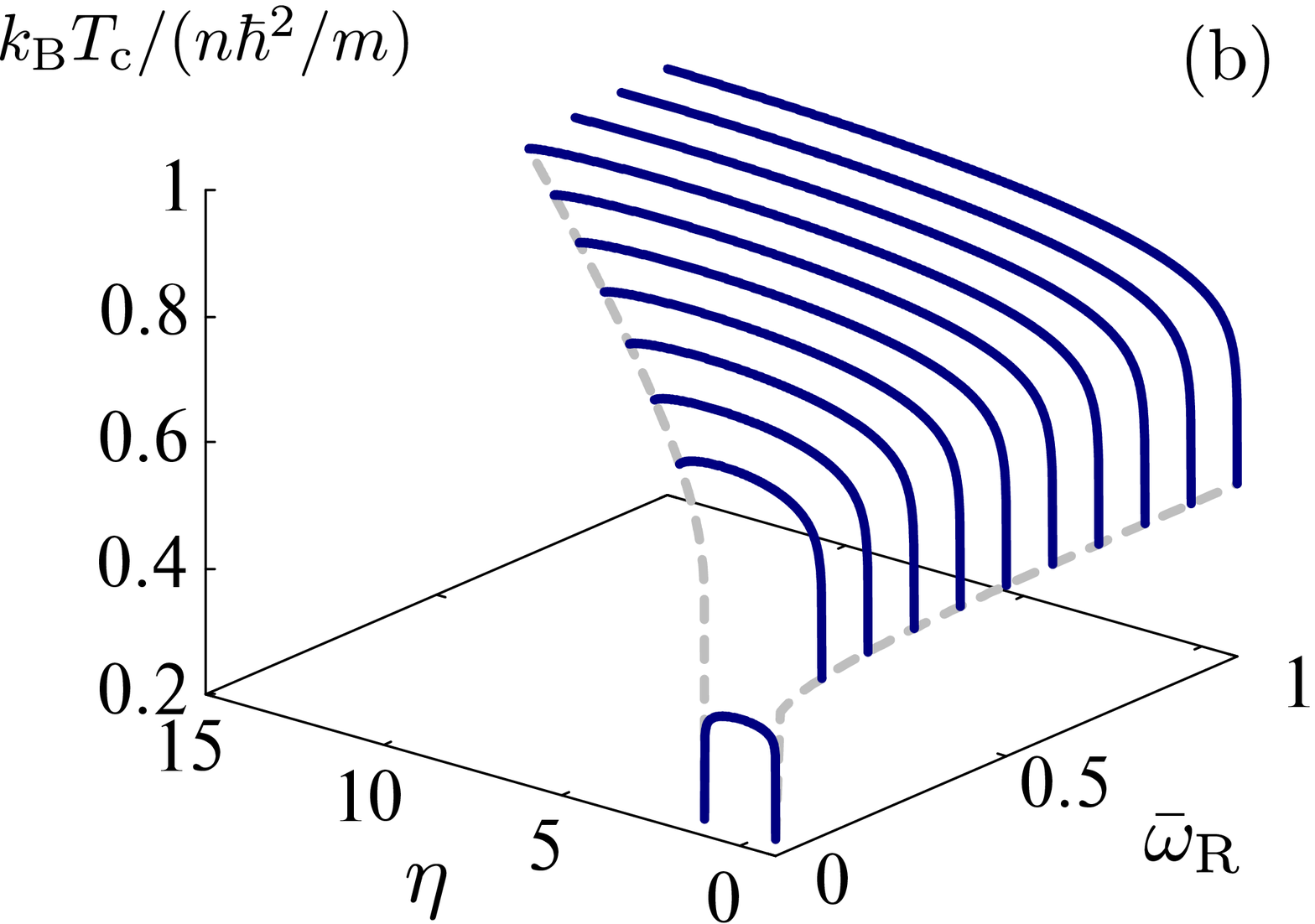}
\end{minipage}
\begin{minipage}[b]{0.49\linewidth}
\centering
\includegraphics[keepaspectratio,scale=0.18]{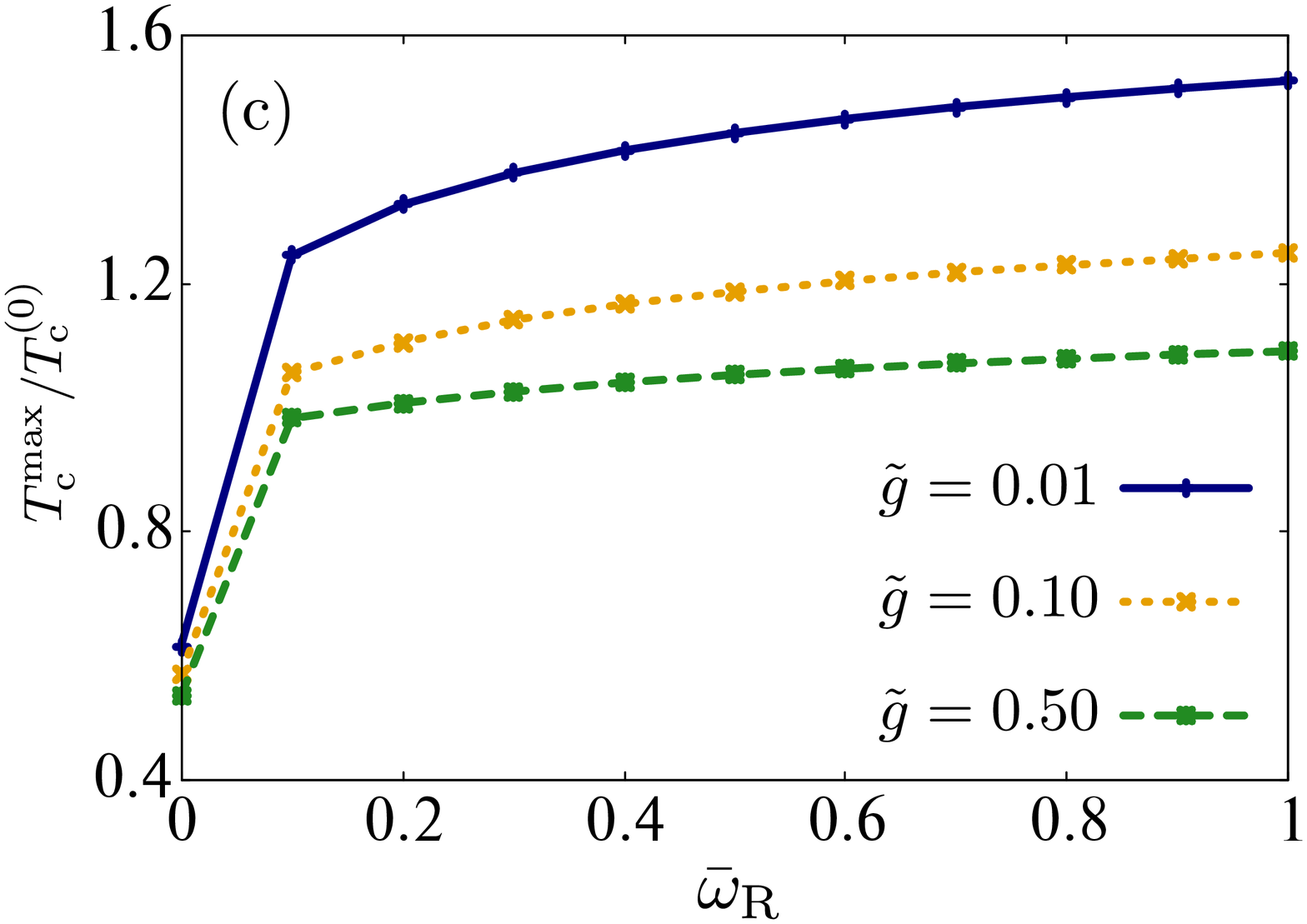}
\end{minipage}
\caption{Phase diagram of the binary Bose mixture and the BKT transition temperature to intercomponent coupling $\eta$ and Rabi coupling $\bar{\omega}_{\mathrm{R}}$. 
The curves in (a) represent the BKT transition temperature for $\Tilde{g}=0.1$ and $\bar{\omega}_{\mathrm{R}}=0.0,0.1,0.5$, below which the system is superfluid (SF). 
Above the transition temperature, it turns into a normal (N) phase with the vanishing superfluid fraction. 
The gray dotted curve in (a) represents the boundary at $\eta=1+2\hbar\omega_{\mathrm{R}}/(gn)$. 
The vertical thin lines represent $\eta=1+2\hbar\omega_{\mathrm{R}}/(gn)$ for each Rabi coupling above which the population-balanced ground state changes to the polarized phase. 
For $\eta<-1$, the population-balanced ground state is unstable. 
The two dashed curves in (b) represent the boundaries of the stable region of the ground state with balanced densities at $\eta=-1$ and $\eta=1+2\hbar\omega_{\mathrm{R}}/(gn)$, respectively. 
(c) shows the maximum value of the BKT transition temperature scaled by the transition temperature in the single-component case $T_{\rm c}^{\text{max}}/T_{\rm c}^{(0)}$, with $\Tilde{g}=0.01,0.1,0.5$. }
\label{FigTBKTrg3deta}
\end{figure}

Figure \ref{FigTBKTrg3deta} shows the phase diagram and the BKT transition temperature. 
In Fig.~\ref{FigTBKTrg3deta}(a), the curves represent the $\eta$ dependence of the BKT transition temperature in the thermodynamic limit with $\Tilde{g}=0.1$ and $\bar{\omega}_{\rm R}=0,0.1,0.5$. 
The shaded region below the transition temperature is the superfluid phase with a finite superfluid density for each of the values of Rabi coupling, while the system is in the normal phase above that temperature. 
We can observe that, as $\eta$ increases from $-1$, the transition temperature first increases. 
Near $\eta=1+2\hbar\omega_{\mathrm{R}}/(gn)$, it reaches a maximum for each $\bar{\omega}_{\mathrm{R}}$ and changes to a gradual decrease. 
In particular, at $\bar{\omega}_{\mathrm{R}}=0$, as displayed in Fig.~\ref{FigTBKTrg3deta}(a), the BKT transition temperature is symmetric with respect to $\eta$ and reaches its maximum at $\eta=0$. 
This is a natural consequence of the two symmetric excitation spectra $E^{(\pm)}_{k}=\sqrt{\varepsilon_{k}\left[\varepsilon_{k}+gn\left(1\mp\eta\right)\right]}$ for $\omega_{\mathrm{R}}=0$. 
Figure \ref{FigTBKTrg3deta}(b) displays a 3D plot of the BKT transition temperature as a function of $\eta$ and $\bar{\omega}_{\mathrm{R}}$. 
It shows the monotonic increase of the transition temperature with increasing Rabi coupling $\bar{\omega}_{\rm R}$. 
This behavior can be explained by the behavior of the energy gap in $E^{(+)}_{k}$ due to the Rabi coupling. 
As one increases the Rabi coupling, the gap size also increases, and the normal density $n_{\rm n}^{(+)}$ in Eq.~\eqref{nnplusminus} decreases, while $n_{\rm n}^{(-)}$ is unaffected. 
This results in an increase of the superfluid density in Eq.~\eqref{ns0}, thereby leading to an enhancement of the BKT transition temperature according to Eq.~\eqref{halfcriterion} by replacing the renormalized superfluid density with the bare one, which is a good approximation at low temperatures as illustrated in Fig.~\ref{Fignsrg0e1}(a). 
The maximum value of the transition temperature scaled by the one in the single-component case is shown in Fig.~\ref{FigTBKTrg3deta}(c) with varying Rabi coupling. 
It monotonically increases by increasing $\bar{\omega}_{\rm R}$. 
Figure \ref{FigTBKTrg3deta}(c) also reveals that the ratio $T_{\rm c}^{\rm max}/T_{\rm c}^{(0)}$ is prominently enhanced as one decreases the intra-coupling strength $\Tilde{g}$. 
This behavior comes from monotonically increasing the critical temperature $T_{\rm c}^{(0)}$ in the single-component Bose gas faster than $T_{\rm c}^{\rm max}$ by increasing $\Tilde{g}$.

\begin{figure}[t]
\begin{minipage}[b]{0.47\linewidth}
\centering
\includegraphics[keepaspectratio,scale=0.18]{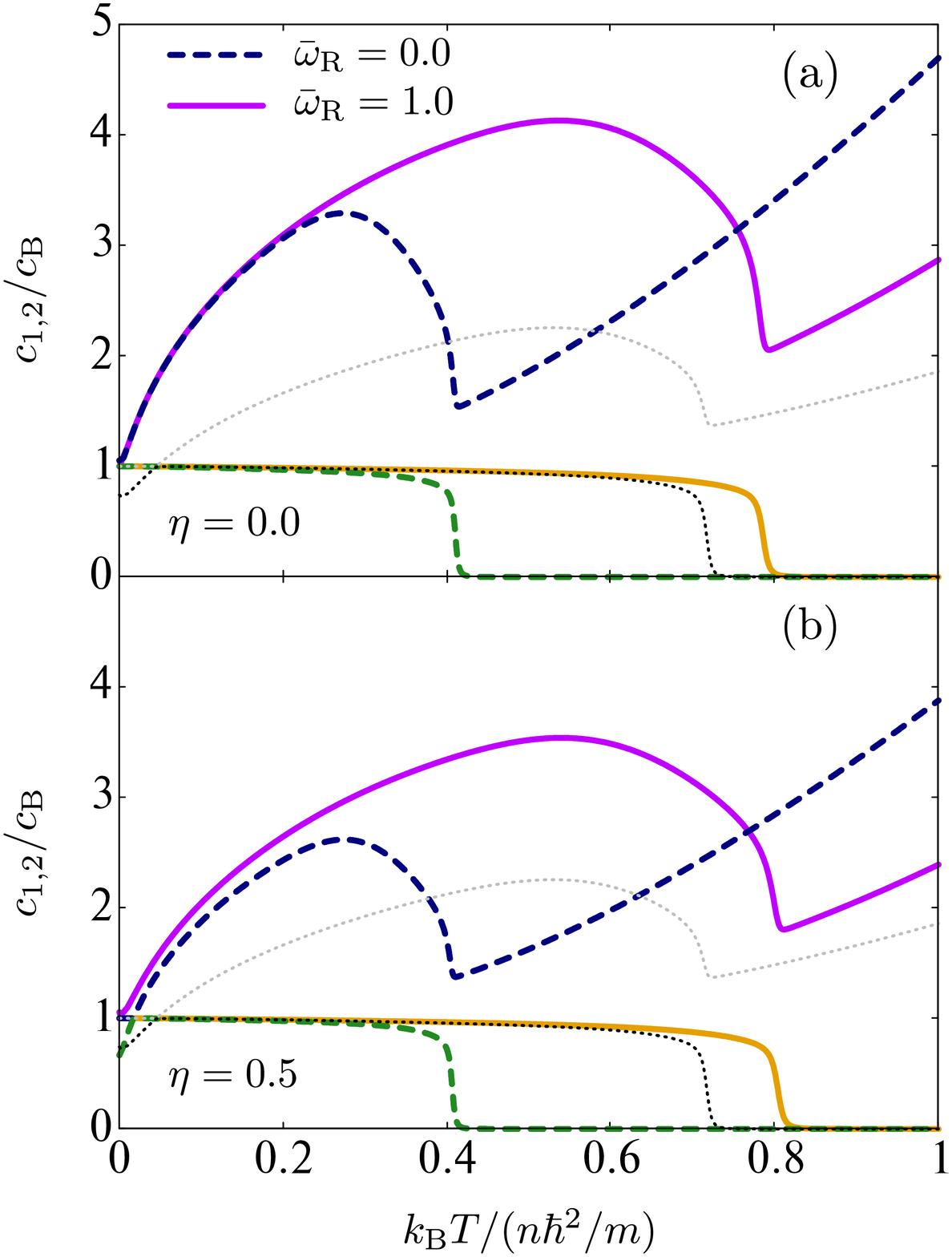}
\end{minipage}
\begin{minipage}[b]{0.47\linewidth}
\centering
\includegraphics[keepaspectratio,scale=0.18]{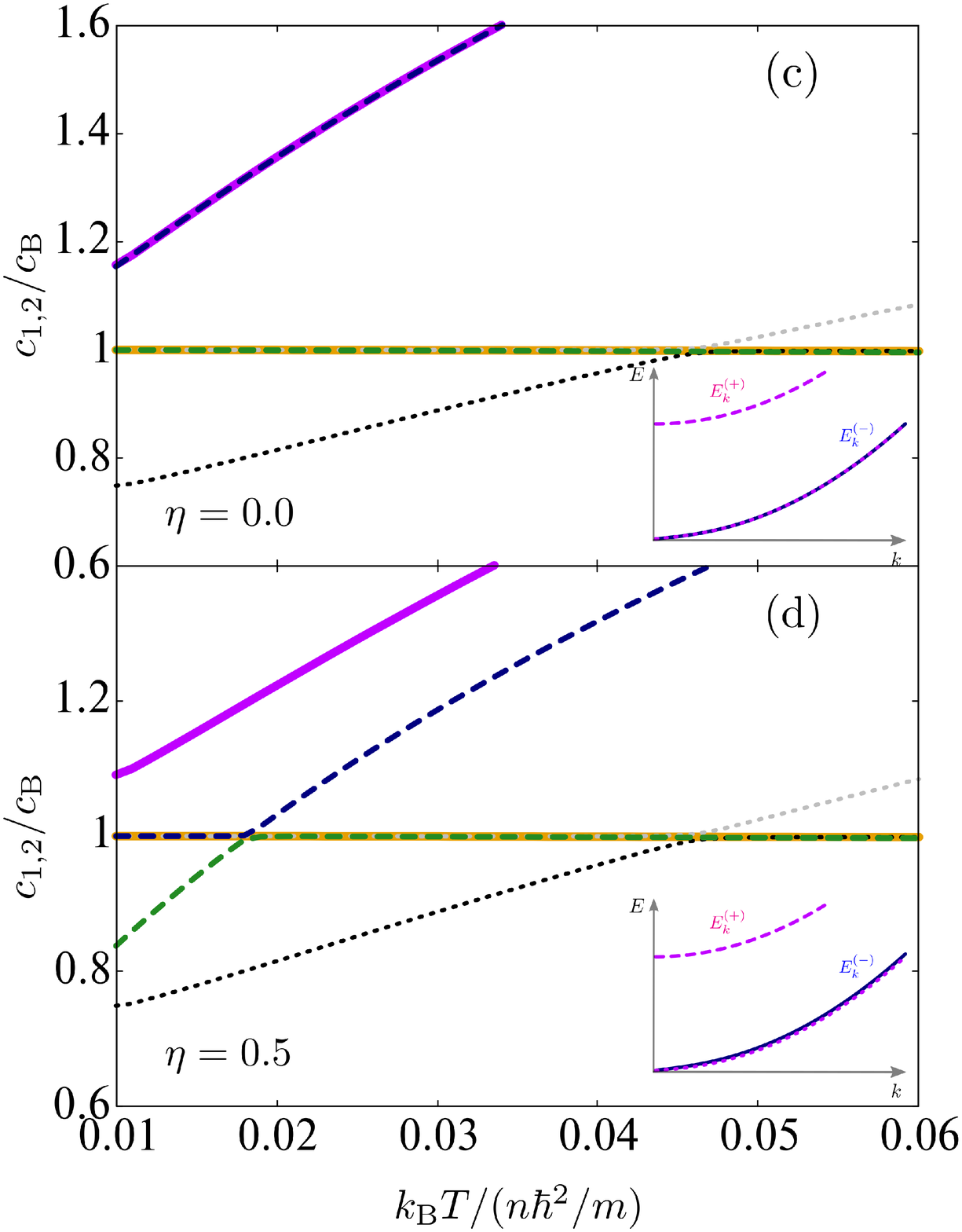}
\end{minipage}
\caption{First and second sound velocities $c_{1,2}$ scaled by the Bogoliubov velocity $c_{\rm B}$ for $\Tilde{g}=0.1$ and $L/\xi=200$. 
The intercoupling is set to be $\eta=0$ in (a) and $\eta=0.5$ in (b). 
The dashed curves correspond to $\bar{\omega}_{\rm R}=0.0$, while the solid curves correspond to $\bar{\omega}_{\rm R}=1.0$. 
The thin dotted curves represent $c_{1,2}$ in a single-component Bose gas for $\Tilde{g}=0.1$. 
The low-temperature behavior is magnified in (c) and (d). 
The insets in (c) and (d) illustrate the elementary excitations $E_{k}^{(\pm)}$. 
The solid curves stand for $E_{k}^{(-)}$, while the dotted and dashed curves represent $E_{k}^{(+)}$ for $\bar{\omega}_{\rm R}=0.0$ and $\bar{\omega}_{\rm R}=1.0$, respectively. }
\label{Figsoundv}
\end{figure}

The propagation of sound waves occurs in a fluid due to density fluctuations, and the sound velocity is determined by thermodynamic properties. 
In a superfluid, in addition to the density wave, there is another collective mode associated with the entropy fluctuations originating from the no-entropy flow in superfluids. 
The collective mode of the entropy wave is called the second sound \cite{landau1941,landaufluid,khala,furutani,stringari}. 
The first and second sound velocities $c_{1,2}$ are the roots of Landau's two-fluid equation $c^{4}-(v_{s}^{2}+v_{\rm L}^{2})c^{2}+v_{T}^{2}v_{\rm L}^{2}=0$, where $v_{T}$, $v_{s}$, and $v_{\rm L}$ are the isothermal, adiabatic, and Landau velocities, respectively, calculated from the free energy \cite{furutani,supplement}. 
Figure \ref{Figsoundv} illustrates the first and second sound velocities for $\Tilde{g}=0.1$ and $\eta=0,0.5$ with $\bar{\omega}_{\rm R}=0.0,1.0$. 
The upper branch is the first sound velocity $c_{1}$, and the lower branch is identified as the second sound velocity $c_{2}$, which survives as long as the superfluid fraction is finite. 
Finite Rabi coupling increases the critical temperature, as shown in Fig.~\ref{Fignsrg0e1}, and allows the second sound to be present up to a higher temperature. 
At the low-temperature limit in the absence of Rabi coupling, using the linear dispersions $E_{k}^{(+)}\simeq c_{+}\hbar k$, with $c_{+}=[(1-\eta)gn/(2m)]^{1/2}$, and $E_{k}^{(-)}\simeq c_{\rm B}\hbar k$, one finds $v_{T}=v_{s}=c_{\rm B}$ and $v_{\rm L}=[(c_{+}^{-2}+c_{\rm B}^{-2})/(c_{+}^{-4}+c_{\rm B}^{-4})]^{1/2}=[(1-\eta)/(1+\eta^{2})]^{1/2}c_{\rm B}$ \cite{supplement}. 
For $\eta=0$ as shown in Fig.~\ref{Figsoundv}(a), in particular, the first and second sound velocities coincide with each other, $c_{1}=c_{2}=v_{T}=v_{s}=v_{\rm L}=c_{\rm B}$. 
The low-temperature behavior is shown in Fig.~\ref{Figsoundv}(c). 
With $0<\eta<1$ in the low-temperature regime without Rabi coupling, one observes $c_{1}=v_{s}=v_{T}=c_{\rm B}$ and $c_{2}=v_{\rm L}<c_{\rm B}$, indicating that the sound modes are identified as the density mode and entropy mode, respectively, as illustrated by the dashed curves in Fig.~\ref{Figsoundv}(d). 
As one increases the temperature, the two branches exhibit a quasicrossing at which the density mode and entropy mode start to mix as in the case of the single-component 2D Bose gas plotted with the thin dotted curves in Fig.~\ref{Figsoundv} or a 3D Bose gas \cite{furutani,stringari}. 
In contrast, the solid curves in Fig.~\ref{Figsoundv} imply that finite Rabi coupling suppresses the quasicrossing, as shown in Fig.~\ref{Figsoundv}(d), which is distinct from a single-component 2D Bose gas. 
This behavior can be understood by the presence of a gapped mode. 
With finite Rabi coupling, $E_{k}^{(+)}$ is gapped out, as shown in the insets in Figs.~\ref{Figsoundv}(c) and \ref{Figsoundv}(d), and most thermally excited bosons occupy only the gapless mode $E_{k}^{(-)}\simeq c_{\rm B}\hbar k$. 
Then, the major difference from the single-component case is only the additional prefactor $1/2$ in Eqs.~\eqref{nnplusminus} which affects the Landau velocity. 
Consequently, the Landau velocity is found to be identical to the Bogoliubov velocity, which also coincides with the adiabatic velocity at zero temperature \cite{supplement}. 
It results in the suppression of quasicrossing at a low temperature. 
The temperature at which the quasicrossing occurs characterizes the temperature above which the second sound can be detected by a density probe \cite{tononi,meppelink,hu,miki}. 
From an experimental point of view, the suppression of quasicrossing at finite temperature implies that the second sound mode is sensitive to a density probe even in the low-temperature regime, which can be tested with ultracold-atom experiments \cite{meppelink,shin2020}.

In summary, we investigated BKT transition in a Rabi-coupled binary Bose mixture under balanced densities. 
We have derived the NK RG equations for a binary Bose mixture and pointed out that the NK criterion is subject to change due to the fractional parameter and the Rabi coupling, consistent with the Monte Carlo simulation \cite{nitta}. 
Based on the obtained RG equations, we clarified the whole behavior of the BKT transition temperature with respect to the Rabi coupling and intercomponent coupling. 
We found a nonmonotonic behavior of the transition temperature in terms of the intercomponent coupling and showed the maximum transition temperature for each value of Rabi coupling finding regimes of parameters resulting in an amplification of the transition temperature. 
Finally, we have studied the first and second sound velocities in this binary Bose mixture. 
We confirmed the jump in the second sound velocity as well as the superfluid density at the BKT transition temperature and elucidated the quasicrossing behavior of the two sound modes in the low-temperature regime. 
Our obtained NK criterion is consistent with the prediction based on Monte Carlo analysis for the population-balanced case \cite{nitta}. 
On the other hand, Monte Carlo analysis has also predicted a double-step structure of the superfluid density in the population-imbalanced case \cite{nitta,abad2013,lellouch2013,search2001,Tommasini2003}. 
A challenging open problem is to obtain a consistent result through the RG analysis in this population-imbalanced Bose mixture, extending the approach investigated in this work \cite{kobayashi2020}. 

\begin{acknowledgments}
The authors thank M. Kobayashi and T. Enss for the useful comments. 
K.F. acknowledges the Ph.D. fellowship of the Fondazione Cassa di Risparmio di Padova e Rovigo. 
\end{acknowledgments}

\widetext
\pagebreak

\renewcommand{\theequation}{S\arabic{equation}}
\renewcommand{\thefigure}{S\arabic{figure}}
\setcounter{equation}{0}
\setcounter{figure}{0}

\begin{center}
\textbf{\large Supplemental Material for "Berezinskii-Kosterlitz-Thouless phase transition with Rabi-coupled bosons"}
%\begin{abstract}
%We provide detailed derivations of the Nelson-Kosterlitz renormalization group equations in a single-component Bose gas and in a binary Bose mixture discussed in the main text. We show that Rabi coupling modifies the renormalization group equations resulting in the recovery of the Nelson-Kosterlitz criterion in a single-component Bose gas. We also describe the details of the calculation of the thermodynamic quantities and sound velocities in the low-temperature regime in a single-component Bose gas and in a two-component Bose mixture. 
%\end{abstract}
\end{center}

\subsection{Derivation of Nelson-Kosterlitz renormalization group equations}

A single-component weakly-interacting Bose gas is described by the Lagrangian density 
\beq
\mathcal{L}=i\hbar\psi^{*}\partial_{t}\psi-\frac{\hbar^{2}}{2m}\abs{\grad\psi}^{2}-\frac{g}{2}\abs{\psi}^{4} ,
\label{Lsingle}
\eeq
with $\psi(\bm{r},t)$ the complex bosonic field, $m$ the atomic mass, and $g$ the interaction strength. 
To address the BKT physics, we employ Popov's treatment: $\psi(\bm{r},t)=\Tilde{\psi}(\bm{r},t)e^{i\theta(\bm{r})}$ assuming a time-independent phase $\theta(\bm{r})$ as a slowly varying field and $\Tilde{\psi}(\bm{r},t)$ as a fast field of the superfluid phase \cite{SMpopov,SMsvistunov}. 
By integrating out the fast variable $\Tilde{\psi}$, the Euclidean action associated with the formation of vortices is given by 
\beq
S[\theta]
=\hbar\int d^{2}\bm{r}\frac{K}{2}\left(\grad\theta\right)^{2} ,
\label{SXY}
\eeq
where $K=\hbar^{2}n_{\rm s}/(mk_{\rm B}T)=J/(k_{\rm B}T)$ with $J=\hbar^{2}n_{\rm s}/m$ the phase stiffness and $n_{\rm s}$ the superfluid density \cite{SMsvistunov}. 
Practically, Eq.~\eqref{SXY} can be obtained just by regarding the quasicondensate density as a uniform superfluid density $n=|\Tilde{\psi}|^{2}=n_{\rm s}$ and inserting it into Eq.~\eqref{Lsingle}. 
In the presence of vortices, the XY model in Eq.~\eqref{SXY}, which is equivalent to a Coulomb gas apart from the analytic spin-wave contribution, can be mapped to the sine-Gordon model described by \cite{SMgiamarchi,SMaltlandsimons}
\beq
S_{\rm sG}[\phi]=\frac{\hbar}{2\pi^{2}K}\int d^{2}\bm{r}\left(\grad\phi\right)^{2}
-\frac{2\hbar y}{\alpha^{2}}\int d^{2}\bm{r}\cos{(2\phi)} ,
\label{SsG}
\eeq
with $\phi(\bm{r})$ the analytic real field for the Coulomb gas, $y\equiv\mathrm{exp}[-\mu_{\mathrm{v}}/(k_{\mathrm{B}}T)]$ the dimensionless parameter characterizing the strength of the cosine potential corresponding to the vortex fugacity where $\mu_{\mathrm{v}}$ is the vortex chemical potential, and $\alpha$ the short-range cutoff \cite{SMgiamarchi,SMaltlandsimons,SMnelson}. 
To develop RG equations, we consider a correlation function \cite{SMgiamarchi}
\beq
R(\bm{r}_{1}-\bm{r}_{2})=\left\langle e^{i\phi(\bm{r}_{1})}e^{-i\phi(\bm{r}_{2})}\right\rangle 
=\frac{1}{Z}\int\mathcal{D}\phi e^{i\phi(\bm{r}_{1})}e^{-i\phi(\bm{r}_{2})} e^{-S_{\rm sG}[\phi]/\hbar}, 
\eeq
where $Z=\int\mathcal{D}\phi\mathrm{exp}\left[-S_{\rm sG}[\phi]/\hbar\right]$ is the partition function. 
Neglecting the cosine potential in Eq.~\eqref{SsG}, we get $R_{0}(\bm{r}_{1}-\bm{r}_{2})=\mathrm{exp}[-\pi KF(\bm{r}_{1}-\bm{r}_{2})/2]$ with $F(\bm{r})=\ln{(\abs{\bm{r}}/\alpha)}$. 
Perturbative expansion in terms of $y$ up to $\order{y^{2}}$ results in
\beq
R(\bm{r}_{1}-\bm{r}_{2})=R_{0}(\bm{r}_{1}-\bm{r}_{2})
\left[1+\frac{y^{2}}{2\alpha^{4}}\sum_{\sigma=\pm1}\int d^{2}\bm{r}'\int d^{2}\bm{r}''e^{-2\pi KF(\bm{r}'-\bm{r}'')}\left[e^{\pi\sigma KG(\bm{r}_{1},\bm{r}_{2};\bm{r}',\bm{r}'')}-1\right]\right] ,
\label{R1}
\eeq
with $G(\bm{r}_{1},\bm{r}_{2};\bm{r}',\bm{r}'')=F(\bm{r}_{1}-\bm{r}')-F(\bm{r}_{1}-\bm{r}'')+F(\bm{r}_{2}-\bm{r}'')-F(\bm{r}_{2}-\bm{r}')$. 
Assuming $\abs{\bm{r}}=\abs{\bm{r}'-\bm{r}''}\ll\abs{\bm{r}'+\bm{r}''}/2$, Eq.~\eqref{R1} reduces to
\beq
R(\bm{r}_{1}-\bm{r}_{2})=R_{0}(\bm{r}_{1}-\bm{r}_{2})
\left[1+\frac{y^{2}}{\pi\alpha^{4}}\frac{\pi^{2}K^{2}}{4}F(\bm{r}_{1}-\bm{r}_{2})\int_{r>\alpha}d^{2}\bm{r}r^{2}e^{-2\pi KF(\bm{r})}\right] .
\eeq
Defining the effective strength $K_{\rm eff}$ by $R(\bm{r})=\mathrm{exp}[-\pi K_{\rm eff}F(\bm{r})/2]$, it is given by
\beq
K_{\rm eff}^{-1}=K^{-1}+4\pi^{3} y^{2}\int_{1}^{\infty}dxx^{3-2\pi K} ,
\label{Keff}
\eeq
with a dimensionless length scale $x=r/\alpha$ up to $\order{y^{2}}$. 
Splitting the spatial integral at a boundary $b=e^{dl}=1+dl$ and introducing
\beq
\Tilde{K}^{-1}=K^{-1}+4\pi^{3} y^{2}\int_{1}^{b}dxx^{3-2\pi K}, \quad
\Tilde{y}=yb^{2-\pi K},
\label{tilde}
\eeq
one obtains 
\beq
K_{\rm eff}^{-1}=\Tilde{K}^{-1}+4\pi^{3} \Tilde{y}^{2}\int_{1}^{\infty}dxx^{3-4\pi \Tilde{K}},
\eeq
which is equivalent to Eq.~\eqref{Keff} after rescaling $x\to x/b$. 
The set of equations \eqref{tilde} leads to the NK RG equations (8) with $K(l)=\hbar^{2}n_{\mathrm{s}}^{(l)}/(mk_{\mathrm{B}}T)$ and $y(l)=\mathrm{exp}[-\mu_{\mathrm{v}}(l)/(k_{\mathrm{B}}T)]$ where $\mu_{\mathrm{v}}(l)$ is the vortex chemical potential at the dimensionless scale $l$ \cite{SMgiamarchi,SMaltlandsimons,SMnelson}. 
The BKT critical temperature $T_{\rm c}^{(0)}$ can be obtained by the NK criterion (9) which provides a fixed point of Eqs.~(8) \cite{SMnelson,SMkosterlitz}. 

A binary Bose mixture is described by
\beq
\mathcal{L}=\sum_{a=1,2}\left[i\hbar\psi_{a}^{*}\partial_{t}\psi_{a}-\frac{\hbar^{2}}{2m}\abs{\grad\psi_{a}}^{2}-\frac{g}{2}\abs{\psi_{a}}^{4}\right]
-g_{12}\abs{\psi_{1}}^{2}\abs{\psi_{2}}^{2}+\hbar\omega_{\rm R}\left[\psi_{1}^{*}\psi_{2}+\psi_{2}^{*}\psi_{1}\right] .
\eeq
With the transformation $\psi_{a=1,2}(\bm{r},t)=\Tilde{\psi}_{a}(\bm{r},t)e^{i\theta_{a}(\bm{r})}$, by integrating out $\Tilde{\psi}_{a}$, the Euclidean action of a binary Bose mixture relevant to the formation of vortices is given by
\beq
S[\theta_{1},\theta_{2}]
=\hbar\int d^{2}\bm{r}\left[\frac{K_{1}}{2}\left(\grad\theta_{1}\right)^{2}+\frac{K_{2}}{2}\left(\grad\theta_{2}\right)^{2}\right] ,
\label{SbiXY}
\eeq
in the absence of Rabi coupling $\omega_{\rm R}=0$ with $K_{a=1,2}=\alpha_{a}K=J_{a}/(k_{\rm B}T)$ where $J_{a}=\alpha_{a}J$ is the phase stiffness of each component with $\alpha_{a}=n_{a}/n$ the fractional parameter. 
As in the single-component XY model, the binary XY model \eqref{SbiXY} can be mapped to the binary sine-Gordon model
\beq
S_{\rm sG}[\phi_{1},\phi_{2}]=\frac{\hbar}{2\pi^{2}}\int d^{2}\bm{r}\left[\frac{1}{K_{1}}\left(\grad\phi_{1}\right)^{2}+\frac{1}{K_{2}}\left(\grad\phi_{2}\right)^{2}\right]
+\frac{2\hbar}{\alpha^{2}}\int d^{2}\bm{r}\left[y_{1}\cos{(2\phi_{1})}+y_{2}\cos{(2\phi_{2})}\right] .
\eeq
We can follow a similar manner in the single-component case to derive the RG equations.  Up to $\order{y_{1}^{2},y_{2}^{2}}$, for the symmetric case $\alpha_{1}=\alpha_{2}=1/2$ with $y_{1}=y_{2}=y$, it provides a set of NK RG equations \cite{SMgiamarchi,SMaltlandsimons,SMenss}
\beq
\partial_{l}K(l)^{-1}=2\pi^{3}y_{a}^{2}, \quad
\partial_{l}y(l)=\left[2-\frac{\pi}{2}K(l)\right]y(l), 
\label{halfRGeqs1}
\eeq
which are Eqs.~(10) for $\omega_{\rm R}=0$. 

In the presence of Rabi coupling, the two components are coupled as
\beq
\mathcal{L}_{\theta}=-\frac{J_{1}}{2}\left(\grad\theta_{1}\right)^{2}-\frac{J_{2}}{2}\left(\grad\theta_{2}\right)^{2}+2\hbar\omega_{\rm R}\sqrt{n_{1}n_{2}}\cos{\left(\theta_{1}-\theta_{2}\right)} .
\label{LbiXYrabi}
\eeq
It provides the equations of motion 
\beq
J_{1}\grad^{2}\theta_{1}=2\hbar\omega_{\rm R}\sin{\left(\theta_{1}-\theta_{2}\right)}, \quad
J_{2}\grad^{2}\theta_{2}=-2\hbar\omega_{\rm R}\sin{\left(\theta_{1}-\theta_{2}\right)}.
\label{eomrabi}
\eeq 
This set of equations of motion leads to domain wall solutions which are metastable states with a finite Rabi coupling \cite{SMson2002}. 
Inserting the equations of motion \eqref{eomrabi} into the Lagrangian \eqref{LbiXYrabi}, we obtain the optimized Lagrangian
\beq
\mathcal{L}^{\rm opt}_{\theta}=-\frac{2J_{1}}{2}\left(\grad\theta_{1}\right)^{2}+\text{const.}+\order{\grad^{4}}, 
\label{Lopt}
\eeq
for $n_{1}=n_{2}=n/2$. 
The action associated with the optimized Lagrangian \eqref{Lopt} is equivalent to Eq.~\eqref{SXY} in the single-component case because $2J_{1}=J$. 
Note that the effective phase stiffness as a coefficient of $-(\grad\theta_{1})^{2}/2$ in Eq.~\eqref{Lopt} is replaced with $J_{1}$ instead of $2J_{1}$ in the absence of Rabi coupling $\omega_{\rm R}=0$ because the two phases are no longer coupled. 
This is consistent with Eqs.~\eqref{halfRGeqs1}. 
As a result, neglecting the higher-order derivatives of $\order{\grad^{4}}$ and following the procedure mentioned above, we restore the RG equations identical to the ones in the single-component case. 
Taking into account Eqs.~\eqref{halfRGeqs1} in the absence of Rabi coupling, we can write the RG equations as in Eqs.~(10). 

\subsection{Finite size effect on the superfluid density}

\begin{figure*}[t]
\centering
\includegraphics[width=70mm]{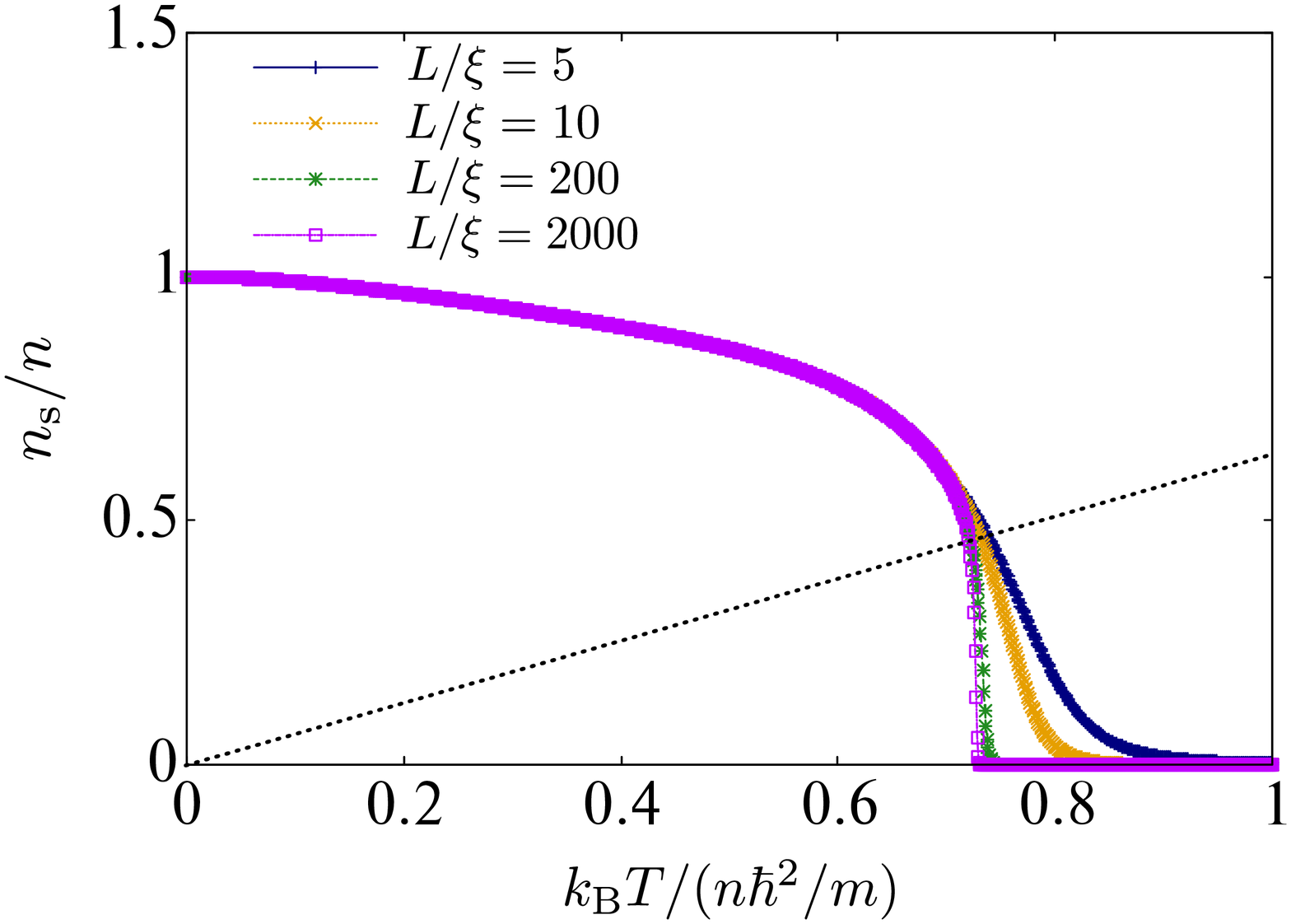}
\caption{Finite size effects on the renormalized superfluid fraction with $\bar{\omega}_{\rm R}=0.1$, $\eta=0.0$, and $\Tilde{g}=0.1$. 
The black dotted thin line represents $k_{\rm B}T=\pi\hbar^{2}n_{\rm s}(T)/(2m)$. }
\label{FigSMnssize}
\end{figure*}

Figure \ref{FigSMnssize} shows the renormalized superfluid fraction with varying system size length $L$ in a Rabi-coupled binary Bose mixture. 
In the thermodynamic limit $L\to\infty$, the renormalized superfluid density is finite below the BKT transition temperature, while it discontinuously drops to zero at the BKT transition temperature, which is determined by the NK criterion (11). 
As mentioned in the main text, Fig.~\ref{FigSMnssize} indicates that a finite system size smears the discontinuous drop. 

\subsection{Thermodynamics and sound velocities}

The isothermal, adiabatic, and Landau velocities are given by
\beq
v_{T}=\sqrt{\frac{1}{m}\left(\frac{\partial P}{\partial n}\right)_{T}}, \quad
v_{s}=\sqrt{\frac{1}{m}\left(\frac{\partial P}{\partial n}\right)_{s}}, \quad
v_{\rm L}=\sqrt{\frac{n_{\rm s}Ts^{2}}{n_{\rm n}c_{V}}},
\label{vTsL}
\eeq
respectively with 
\beq
P=-\left(\frac{\partial F}{\partial L^{2}}\right)_{N,T} ,\quad
s=\frac{1}{mN}\left(\frac{\partial F}{\partial T}\right)_{N,L^{2}} ,\quad
c_{V}=T\left(\frac{\partial s}{\partial T}\right)_{N,L^{2}} ,
\label{Psc}
\eeq
the pressure, the entropy per mass unit, and the specific heat at constant volume respectively. 
In a 2D single-component Bose gas, the free energy $F$ at the mean-field level is given by \cite{SMfurutani}
\beq
F=\frac{gN^{2}}{2L^{2}} + L^{2}k_{\rm B}T\int\frac{d^{2}\bm{k}}{(2\pi)^{2}}\ln{\left[1-e^{-E_{k}/(k_{\rm B}T)}\right]} ,
\eeq
with the Bogoliubov spectrum $E_{k}=\sqrt{\varepsilon_{k}(\varepsilon_{k}+2gn)}$. 
The first term is free energy at zero temperature. 
The second term involving thermal excitations represents the thermal contribution at a finite temperature. 
In the phononic regime at low temperatures, the linear dispersion $E_{k}=c_{\rm B}\hbar k$ with $c_{\rm B}=\sqrt{gn/m}$ gives analytic expressions of the free energy and the normal density as
\beq
\frac{F}{N}=\frac{gn}{2}-\frac{\zeta(3)}{2\pi\hbar^{2}}\frac{(k_{\rm B}T)^{3}}{nc_{\rm B}^{2}} ,\quad
n_{\rm n}=-\int{d^2{\bm k}\over (2\pi)^2} {\hbar^2k^2\over 2m} f_{T}'(E_{k})
=\frac{3\zeta(3)}{2\pi\hbar^{2}}\frac{(k_{\rm B}T)^{3}}{mc_{\rm B}^{4}} ,
\label{singleFns}
\eeq
with $\zeta(x)$ the Riemann zeta function \cite{SMfurutani,SMlandau1941}. 
At zero temperature, Eqs.~\eqref{vTsL} with Eqs.~\eqref{Psc} and \eqref{singleFns} lead to 
\beq
c_{1}=v_{T}=v_{s}=c_{\rm B}, \quad
c_{2}=v_{\rm L}=\frac{c_{\rm B}}{\sqrt{2}} .
\label{c12single}
\eeq

In a two-component Bose mixture, the free energy $F$ at the mean-field level is given by
\beq
F=\frac{1+\eta}{4}\frac{gN^{2}}{L^{2}}-\hbar\omega_{\rm R}N 
+L^{2}k_{\rm B}T\int\frac{d^{2}\bm{k}}{(2\pi)^{2}}
\Big[\ln{\left(1-e^{-E_{k}^{(-)}/(k_{\rm B}T)}\right)} 
+\ln{\left(1-e^{-E_{k}^{(+)}/(k_{\rm B}T)}\right)}\Big] .
\label{Fbi}
\eeq
In the phononic regime at low temperatures without Rabi coupling $\omega_{\rm R}=0$, $E_{k}^{(-)}=c_{\rm B}\hbar k$ with $c_{\rm B}=\sqrt{(1+\eta)gn/(2m)}$ and $E_{k}^{(+)}=c_{+}\hbar k$ with $c_{+}=\sqrt{(1-\eta)gn/(2m)}$ provide 
\beq
\frac{F}{N}=\frac{1+\eta}{2}\frac{gn}{2}-\hbar\omega_{\rm R}-\frac{\zeta(3)}{2\pi\hbar^{2}}\frac{(k_{\rm B}T)^{3}}{n}\left(\frac{1}{c_{+}^{2}}+\frac{1}{c_{\rm B}^{2}}\right) ,\quad
n_{\rm n}=n_{\rm n}^{(+)}+n_{\rm n}^{(-)}
=\frac{3\zeta(3)}{2\pi\hbar^{2}}\frac{(k_{\rm B}T)^{3}}{2m}\left(\frac{1}{c_{+}^{4}}+\frac{1}{c_{\rm B}^{4}}\right) , 
\label{biphFns}
\eeq
which results in
\beq
v_{T}=v_{s}=c_{\rm B}, \quad
v_{\rm L}=\sqrt{\frac{c_{+}^{-2}+c_{\rm B}^{-2}}{c_{+}^{-4}+c_{\rm B}^{-4}}}
=\sqrt{\frac{1-\eta}{1+\eta^{2}}}c_{\rm B} ,
\label{c12bizero}
\eeq
at zero temperature. 
For $-1<\eta\le0$, $v_{\rm L}\ge v_{s}$ leads to $c_{1}=v_{\rm L}$ and $c_{2}=v_{s}$. 
For $0<\eta<1$, $v_{\rm L}<v_{s}$ leads to $c_{1}=v_{s}$ and $c_{2}=v_{\rm L}$. 
With a finite Rabi coupling $\omega_{\rm R}>0$, on the other hand, the thermal contribution associated with the gapped branch $E_{k}^{(+)}$ vanishes at zero temperature. 
Then, the free energy and normal density are given by
\beq
\frac{F}{N}=\frac{1+\eta}{2}\frac{gn}{2}-\hbar\omega_{\rm R}-\frac{\zeta(3)}{2\pi\hbar^{2}}\frac{(k_{\rm B}T)^{3}}{nc_{\rm B}^{2}} ,\quad
n_{\rm n}=\frac{3\zeta(3)}{2\pi\hbar^{2}}\frac{(k_{\rm B}T)^{3}}{2mc_{\rm B}^{4}} . 
\label{biFns}
\eeq
The thermal part of free energy is identical to the one in the single-component case while the normal density is half of that in the single-component case in Eqs.~\eqref{singleFns} because of the prefactor $1/2$ in Eqs.~(7). 
As a result, the sound velocities at zero temperature change to
\beq
c_{1,2}=v_{T}=v_{s}=v_{\rm L}=c_{\rm B} ,
\eeq
at any value of inter-component coupling $\eta$. 
The difference from the single-component case in Eqs.~\eqref{c12single} is ascribed to the modification of the normal density in Eqs.~\eqref{biFns}.


\begin{thebibliography}{999}


\bibitem{kadanoff} J. W. Kane and L. P. Kadanoff, Long-range order in superfluid helium,  Phys. Rev. {\bf 155}, 80 (1967). 

\bibitem{clow} J. R. Clow and J. D. Reppy, Temperature dependence of superfluid critical velocities near $T_{\lambda}$, Phys. Rev. Lett. {\bf 19}, 291 (1967). 

\bibitem{reppy} G. Kukich, R. P. Henkel, and J. D. Reppy, Decay of Superfluid ``Persistent Currents," Phys. Rev. Lett. {\bf 21}, 197 (1968). 

\bibitem{amit} D. J. Amit, Phase transition in HeII films, Phys. Lett. A {\bf 26}, 448 (1968); New form for the thermodynamic potential of He II near $T_{\lambda}$, {\bf 26}, 466 (1968). 

\bibitem{bergman} R. S. Kagiwada, J. C. Fraser, I. Rudnick, and D. Bergman, Superflow in Helium films: third sound measurements, Phys. Rev. Lett. {\bf 22}, 338 (1969). 

\bibitem{henkel} R. P. Henkel, E. N. Smith, and J. D. Reppy, Temperature dependence of the superfluid healing length, Phys. Rev. Lett. {\bf 23}, 1276 (1969).

\bibitem{chester72} M. Chester, L. C. Yang, and J. B. Stephens, Quartz microbalance studies of an adsorbed Helium film, Phys. Rev. Lett. {\bf 29}, 211 (1972). 

\bibitem{chester73} M. Chester and L. C. Yang, Superfluid fraction in thin Helium films, Phys. Rev. Lett. {\bf 31}, 1377 (1973). 

\bibitem{chan} M. H. W. Chan, A. W. Yanof, and J. D. Reppy, Superfluidity of thin $^{4}$He films, Phys. Rev. Lett. {\bf 32}, 1347 (1974). 

\bibitem{berthold} J. E. Berthold, D. J. Bishop, and J. D. Reppy, Superfluid Transition of $^{4}$He Films Adsorbed on Porous Vycor Glass, Phys. Rev. Lett. {\bf 39}, 348 (1977).

\bibitem{bishop78} D. J. Bishop and J. D. Reppy, Study of the Superfluid Transition in Two-Dimensional $^{4}$He Films, Phys. Rev. Lett. {\bf 40}, 1727 (1978). 

\bibitem{bishop80} D. J. Bishop and J. D. Reppy, Study of the superfluid transition in two-dimensional $^{4}$He films, Phys. Rev. B {\bf 22}, 5171 (1980). 

\bibitem{bishop81} D. J. Bishop, J. E. Berthold, J. M. Parpia, and J. D. Reppy, Superfluid density of thin $^{4}$He films adsorbed in porous Vycor glass, 
Phys. Rev. B {\bf 24}, 5047 (1981). 

\bibitem{kotsubo} V. Kotsubo and G. A. Williams, Kosterlitz-Thouless Superfluid Transition for Helium in Packed Powders, 
Phys. Rev. Lett. {\bf 53}, 691 (1984). 

\bibitem{minnhagen} P. Minnhagen, The two-dimensional Coulomb gas, vortex unbinding, and superfluid-superconducting films, Rev. Mod. Phys. {\bf 59}, 1001 (1987). 

\bibitem{agnolet} G. Agnolet, D. F. McQueeney, and J. D. Reppy, Kosterlitz-Thouless transition in helium films, 
Phys. Rev. B {\bf 39}, 8934 (1989). 

\bibitem{hadzibabic2006} Z. Hadzibabic, P. Kr{\"u}ger, M. Cheneau, B. Battelier, and J. Dalibard, Berezinskii-Kosterlitz-Thouless crossover in a trapped atomic gas, Nature {\bf 441}, 1118 (2006). 
 
\bibitem{phillips} P. Clad{\'e}, C. Ryu, A. Ramanathan, K. Helmerson, and W. D. Phillips, Observation of a 2D Bose gas: from thermal to quasicondensate to superfluid, 
Phys. Rev. Lett. {\bf 102}, 170401 (2009). 

\bibitem{cornell} S. Tung, G. Lamporesi, D. Lobser, L. Xia, and E. A. Cornell, Observation of the presuperfluid regime in a two-dimensional Bose gas, 
Phys. Rev. Lett. {\bf 105}, 230408 (2010). 

\bibitem{turlapov} K. Martiyanov, V. Makhalov, and A. Turlapov, Observation of a two-dimensional Fermi gas of atoms, Phys. Rev. Lett. {\bf 105}, 030404 (2010).  

\bibitem{dyke} P. Dyke, E. D. Kuhnle, S. Whitlock, H. Hu, M. Mark, S. Hoinka, M. Lingham, P. Hannaford, and C. J. Vale, Crossover from 2D to 3D in a weakly interacting Fermi gas, Phys. Rev. Lett. {\bf 106}, 105304 (2011). 

\bibitem{dalibard2011} T. Yefsah, R. Desbuquois, L. Chomaz, K. J. G{\"u}nter, and J. Dalibard, Exploring the Thermodynamics of a Two-Dimensional Bose Gas, 
Phys. Rev. Lett. {\bf 107}, 130401 (2011). 

\bibitem{chin2011} C.-L. Hung, X. Zhang, N. Gemelke, and C. Chin, Observation of scale invariance and universality in two-dimensional Bose gases, 
Nature {\bf 470}, 236 (2011). 

\bibitem{feld} M. Feld, B. Fr{\"o}hlich, E. Vogt, M. Koschorreck, and M. K{\"o}hl, Observation of a pairing pseudogap in a two-dimensional Fermi gas, Nature {\bf 480}, 75 (2011). 

\bibitem{zwierlein} A. T. Sommer, L. W. Cheuk, M. J. H. Ku, W. S. Bakr, and M. W. Zwierlein, Evolution of fermion pairing from three to two dimensions, Phys. Rev. Lett. {\bf 108}, 045302 (2012). 

\bibitem{kohl} M. Koschorreck, D. Pertot, E. Vogt, B. Fr{\"o}hlich, M. Feld, and M. K{\"o}hl, Attractive and repulsive Fermi polarons in two dimensions, Nature {\bf 485}, 619 (2012). 

\bibitem{vogt} E. Vogt, M. Feld, B. Fr{\"o}hlich, D. Pertot, M. Koschorreck, and M. K{\"o}hl, Scale invariance and viscosity of a two-dimensional Fermi gas, Phys. Rev. Lett. {\bf 108}, 070404 (2012). 

\bibitem{dalibard2012} R. Desbuquois, L. Chomaz, T. Yefsah, J. L{\'e}onard, J. Beugnon, C.  Weitenberg, and J. Dalibard, Superfluid behaviour of a two-dimensional Bose gas,  
Nature Physics {\bf 8}, 645 (2012). 

\bibitem{chin2013} L.-C. Ha, C.-L. Hung, X. Zhang, U. Eismann, S.-K. Tung, and C. Chin, Strongly Interacting Two-Dimensional Bose Gases, 
Phys. Rev. Lett. {\bf 110}, 145302 (2013). 

\bibitem{ries} M. G. Ries, A. N. Wenz, G. Z{\"u}rn, L. Bayha, I. Boettcher, D. Kedar, P. A. Murthy, M. Neidig, T. Lompe, and S. Jochim, Observation of Pair Condensation in the Quasi-2D BEC-BCS Crossover, 
Phys. Rev. Lett. {\bf 114}, 230401 (2015). 

\bibitem{hadzibabic2015} R. J. Fletcher, M. Robert-de-Saint-Vincent, J. Man, N. Navon, R. P. Smith, K. G. H. Viebahn, and Z. Hadzibabic, Connecting Berezinskii-Kosterlitz-Thouless and BEC Phase Transitions by Tuning Interactions in a Trapped Gas, 
Phys. Rev. Lett. {\bf 114}, 255302 (2015). 

\bibitem{murthy} P. A. Murthy, I. Boettcher, L. Bayha, M. Holzmann, D. Kedar, M. Neidig, M. G. Ries, A. N. Wenz, G. Z{\"u}rn, and S. Jochim, Observation of the Berezinskii-Kosterlitz-Thouless phase transition in an ultracold Fermi gas, 
Phys. Rev. Lett. {\bf 115}, 010401 (2015). 

\bibitem{ville} J. L. Ville, R. Saint-Jalm, {\'E}. Le Cerf, M. Aidelsburger, S. Nascimb{\`e}ne, J. Dalibard, and J. Beugnon, Sound Propagation in a Uniform Superfluid Two-Dimensional Bose Gas, 
Phys. Rev. Lett. {\bf 121}, 145301 (2018). 

\bibitem{bohlen} M. Bohlen, L. Sobirey, N. Luick, H. Biss, T. Enss, T. Lompe, and H. Moritz, Sound Propagation and Quantum-Limited Damping in a Two-Dimensional Fermi Gas, 
Phys. Rev. Lett. {\bf 124}, 240403 (2020). 

\bibitem{lompe} L. Sobirey, N. Luick, M. Bohlen, H. Biss, H. Moritz, and T. Lompe, Observation of superfluidity in a strongly correlated two-dimensional Fermi gas, Science, {\bf 372}, 844 (2021). 

\bibitem{hadzibabic2021} P. Christodoulou, M. Ga\l ka, N. Dogra, R. Lopes, J. Schmitt, and Z. Hadzibabic, Observation of first and second sound in a BKT superfluid,
 Nature {\bf 594}, 191 (2021). 
 
\bibitem{pitaevskii} L. Pitaevskii and S. Stringari, {\it Bose-Einstein Condensation and Superfluidity} (Oxford University Press, Oxford, 2016). 

\bibitem{svistunov} B. Svistunov, E. Babaev, and N. Prokof'ev, {\it Superfluid States of Matter} (CRC Press, Boca Raton, FL, 2015). 

\bibitem{tononi19} A. Tononi and L. Salasnich, Bose-Einstein condensation on the surface of a sphere, 
Phys. Rev. Lett. {\bf 123}, 160403 (2019).

\bibitem{tononi20} A. Tononi, F. Cinti, and L. Salasnich, Quantum bubbles in microgravity, 
Phys. Rev. Lett. {\bf 125}, 010402 (2020). 

\bibitem{tononi22} A. Tononi, A. Pelster, and L. Salasnich, Topological superfluid transition in bubble-trapped condensates, 
Phys. Rev. Research {\bf 4}, 013122 (2022). 

\bibitem{lundblad} R. A. Carollo, D. C. Aveline, B. Rhyno, S. Vishveshwara, C. Lannert, J. D. Murphree, E. R. Elliott, J. R. Williams, R. J. Thompson, and N. Lundblad, 
Observation of ultracold atomic bubbles in orbital microgravity, 
Nature {\bf 606}, 281 (2022). 

\bibitem{boyce} R. Y. Chiao and J. Boyce, Bogoliubov dispersion relation and the possibility of superfluidity for weakly interacting photons in a two-dimensional photon fluid, Phys. Rev. A {\bf 60}, 4114 (1999). 

\bibitem{willander} Y. E. Lozovik, A. G. Semenov, and M. Willander, Kosterlitz-Thouless phase transition in microcavity polariton system, JETP Lett. {\bf 84}, 146 (2006). 

\bibitem{brameti} A. Amo, J. Lefr\`{e}re, S. Pigeon, C. Adrados, C. Ciuti, I. Carusotto, R. Houdr{\'e}, E. Giacobino, and A. Bramati, 
Superfluidity of polaritons in semiconductor microcavities, 
Nature Physics {\bf 5}, 805 (2009). 

\bibitem{bloch} D. Sanvitto, F. M. Marchetti, M. H. Szyma{\'n}ska, G. Tosi, M. Baudisch, F. P. Laussy, D. N. Krizhanovskii, M. S. Skolnick, L. Marrucci, A. Lema\^{i}tre, J. Bloch, C. Tejedor, and L. Vi\~{n}a, 
Persistent currents and quantized vortices in a polariton superfluid, 
Nature Physics {\bf 6}, 527 (2010). 

\bibitem{ciuti} I. Carusotto and C. Ciuti, Quantum fluids of light, 
Rev. Mod. Phys. {\bf 85}, 299 (2013). 

\bibitem{yamamoto} W. H. Nitsche, N. Y. Kim, G. Roumpos, C. Schneider, M. Kamp, S. H{\"o}fling, A. Forchel, and Y. Yamamoto, 
Algebraic order and the Berezinskii-Kosterlitz-Thouless transition in an exciton-polariton gas, 
Phys. Rev. B {\bf 90}, 205430 (2014). 

\bibitem{caputo} D. Caputo, D. Ballarini, G. Dagvadorj, C. S{\'a}nchez Mu\~{n}oz, M. De Giorgi, L. Dominici, K. West, L. N. Pfeiffer, G. Gigli, F. P. Laussy, M. H. Szyma{\'n}ska, and D. Sanvitto, 
Topological order and thermal equilibrium in polariton condensates, 
Nat. Mater. {\bf 17}, 145 (2018). 

\bibitem{szymanska} G. Dagvadorj, P. Comaron, and M. H. Szyma{\'n}ska, Unconventional Berezinskii-Kosterlitz-Thouless Transition in the Multicomponent Polariton System, 
arXiv:2208.04167. 

\bibitem{berezinskii} V. L. Berezinskii, Destruction of long-range order in one-dimensional and two-dimensional systems possessing a continuous symmetry group. II. Quantum systems, 
Sov. Phys. JETP {\bf 34}, 610 (1972). 

\bibitem{kosterlitz} J. M. Kosterlitz and D. J. Thouless, Ordering, metastability and phase transitions in two-dimensional systems, 
J. Phys. C {\bf 6}, 1181 (1973). 

\bibitem{nelson} D. R. Nelson and J. M. Kosterlitz, Universal Jump in the Superfluid Density of Two-Dimensional Superfluids, 
Phys. Rev. Lett. {\bf 39}, 1201 (1977). 

\bibitem{kadin} K. Epstein, A. M. Goldman, and A. M. Kadin, Vortex-Antivortex Pair Dissociation in Two-Dimensional Superconductors, 
Phys. Rev. Lett. {\bf 47}, 534 (1981). 

\bibitem{fiory} A. F. Hebard and A. T. Fiory, Critical-Exponent Measurements of a Two-Dimensional Superconductor, 
Phys. Rev. Lett. {\bf 50}, 1603 (1983). 

\bibitem{xue} W. Zhao, Q. Wang, M. Liu, W. Zhang, Y. Wang, M. Chen, Y. Guo, K. He, X. Chen, Y. Wang, J. Wang, X. Xie, Q. Niu, L. Wang, X. Ma, J. K. Jain, M. H. W. Chan, and Q.-K. Xue, 
Evidence for Berezinskii-Kosterlitz-Thouless transition in atomically flat two-dimensional Pb superconducting films, 
Solid State Commun. {\bf 165}, 59 (2013). 

\bibitem{zhao} Z. Lin, C. Mei, L. Wei, Z. Sun, S. Wu, H. Huang, S. Zhang, C. Liu, Y. Feng, H. Tian, H. Yang, J. Li, Y. Wang, G. Zhang, Y. Lu, and Y. Zhao, 
Quasi-two-dimensional superconductivity in $\rm FeSe_{0.3}Te_{0.7}$ thin films and electric-field modulation of superconducting transition, 
Sci. Rep. {\bf 5}, 14133 (2015). 

\bibitem{sharma2022} M. Sharma, M. Singh, R. K. Rakshit, S. P. Singh, M. Fretto, N. De Leo, A. Perali, and N. Pinto, 
Complex Phase-Fluctuation Effects Correlated with Granularity in Superconducting NbN Nanofilms, 
Nanomaterials 12, 4109 (2022).

\bibitem{perali2013} A. Perali, D. Neilson, and A. R. Hamilton, 
High-Temperature Superfluidity in Double-Bilayer Graphene, 
Phys. Rev. Lett. {\bf 110}, 146803 (2013). 

\bibitem{wang2019} Z. Wang, D. A. Rhodes, K. Watanabe, T. Taniguchi, J. C. Hone, J. Shan, and K. F. Mak, 
Evidence of high-temperature exciton condensation in two-dimensional atomic double layers, 
Nature {\bf 574}, 76 (2019). 

\bibitem{mora} C. Mora and Y. Castin, Ground State Energy of the Two-Dimensional Weakly Interacting Bose Gas: First Correction beyond Bogoliubov Theory, 
Phys. Rev. Lett. {\bf 102}, 180404 (2009). 

\bibitem{ozawa} T. Ozawa and S. Stringari, Discontinuities in the First and Second Sound Velocities at the Berezinskii-Kosterlitz-Thouless Transition, 
Phys. Rev. Lett. {\bf 112}, 025302 (2014). 

\bibitem{salasnich2016} L. Salasnich and F. Toigo, Zero-point energy of ultracold atoms, 
Phys. Rep. {\bf 640}, 1 (2016). 

\bibitem{miki} M. Ota and S. Stringari, Second sound in a two-dimensional Bose gas: From the weakly to the strongly interacting regime, 
Phys. Rev. A {\bf 97}, 033604 (2018). 

\bibitem{furutani} K. Furutani, A. Tononi, and L. Salasnich, Sound modes in collisional superfluid Bose gases, 
New. J. Phys. {\bf 23}, 043043 (2021). 

\bibitem{singh} V. P. Singh and L. Mathey, Collective modes and superfluidity of a two-dimensional ultracold Bose gas, 
Phys. Rev. Research {\bf 3}, 023112 (2021). 

\bibitem{singh22} V. P. Singh and L. Mathey, First and second sound in a dilute Bose gas across the BKT transition, 
New. J. Phys. {\bf 24}, 073024 (2022). 

\bibitem{prokofev2001} N. Prokof'ev, O. Ruebenacker, and B. Svistunov, Critical Point of a Weakly Interacting Two-Dimensional Bose Gas, 
Phys. Rev. Lett. {\bf 87}, 270402 (2001). 

\bibitem{prokofev2002} N. Prokof'ev and B. Svistunov, Two-dimensional weakly interacting Bose gas in the fluctuation region, 
Phys. Rev. A {\bf 66}, 043608 (2002). 

\bibitem{dupuis} A. Ran\c{c}on and N. Dupuis, Universal thermodynamics of a two-dimensional Bose gas, 
Phys. Rev. A {\bf 85}, 063607 (2012). 

\bibitem{andreev} A. F. Andreev and E. P. Bashkin, Three-velocity hydrodynamics of superfluid solutions, 
Zh. Eksp. Teor. Fiz. {\bf 69}, 319 (1975) [Sov. Phys. JETP {\bf 42}, 164 (1975)]. 

\bibitem{fil} D. V. Fil and S. I. Shevchenko, Nondissipative drag of superflow in a two-component Bose gas, 
Phys. Rev. A {\bf 72}, 013616 (2005). 

\bibitem{recati2017} J. Nespolo, G. E. Astrakharchik, and A. Recati, Andreev-Bashkin effect in superfluid cold gases mixtures, New J. Phys. {\bf 19}, 125005 (2017). 

\bibitem{konietin} P. Konietin and V. Pastukhov, 2D dilute Bose mixture at low temperatures, 
J. Low Temp. Phys. {\bf 190}, 256 (2018). 

\bibitem{enss} V. Karle, N. Defenu, and T. Enss, Coupled superfluidity of binary Bose mixtures in two dimensions, 
Phys. Rev. A {\bf 99}, 063627 (2019). 

\bibitem{shin2020} J. H. Kim, D. Hong, and Y. Shin, Observation of two sound modes in a binary superfluid gas, 
Phys. Rev. A {\bf 101}, 061601(R) (2020). 

\bibitem{recati2021} D. Romito, C. Lobo, and A. Recati, Linear response study of collisionless spin drag, 
Phys. Rev. Research {\bf 3}, 023196 (2021). 

\bibitem{son2002} D. T. Son and M. A. Stephanov, Domain walls of relative phase in two-component Bose-Einstein condensates, 
Phys. Rev. A. {\bf 65}, 063621 (2002). 

\bibitem{mueller} E. J. Mueller and T.-L. Ho, Two-Component Bose-Einstein Condensates with a Large Number of Vortices, 
Phys. Rev. Lett. {\bf 88}, 180403 (2002). 

\bibitem{kasamatsu2003} K. Kasamatsu, M. Tsubota, and M. Ueda, Vortex Phase Diagram in Rotating Two-Component Bose-Einstein Condensates, 
Phys. Rev. Lett. {\bf 91}, 150406 (2003). 

\bibitem{kasamatsu2004} K. Kasamatsu, M. Tsubota, and M. Ueda, Vortex Molecules in Coherently Coupled Two-Component Bose-Einstein Condensates, 
Phys. Rev. Lett. {\bf 93}, 250406 (2004). 

\bibitem{kasamatsu2005} K. Kasamatsu, M. Tsubota, and M. Ueda, Vortices in Multicomponent Bose-Einstein Condensates, 
Int. J. Mod. Phys. B {\bf 19}, 1835 (2005).

\bibitem{kasamatsu2009} K. Kasamatsu and M. Tsubota, Vortex sheet in rotating two-component Bose-Einstein condensates, 
Phys. Rev. A {\bf 79}, 023606 (2009). 

\bibitem{eto2011} M. Eto, K. Kasamatsu, M. Nitta, H. Takeuchi, and M.Tsubota, Interaction of half-quantized vortices in two-component Bose-Einstein condensates, 
Phys. Rev. A {\bf 83}, 063603 (2011). 

\bibitem{wei} A. Aftalion, P. Mason, and J. Wei, Vortex-peak interaction and lattice shape in rotating two-component Bose-Einstein condensates, 
Phys. Rev. A {\bf 85}, 033614 (2012). 

\bibitem{kuo} P. Kuopanportti, J. A. M. Huhtam{\"a}ki, and M. M{\"o}tt{\"o}nen, Exotic vortex lattices in two-species Bose-Einstein condensates, 
Phys. Rev. A {\bf 85}, 043613 (2012). 

\bibitem{eto2012} M. Eto and M. Nitta, Vortex trimer in three-component Bose-Einstein condensates, 
Phys. Rev. A {\bf 85}, 053645 (2012). 

\bibitem{eto2013} M. Eto and M. Nitta, Vortex graphs as N-omers and $\mathbb{C}P^{N-1}$  skyrmions in N-component Bose-Einstein condensates, 
Europhys. Lett. {\bf 103}, 60006 (2013). 

\bibitem{nitta2013L} M. Cipriani and M. Nitta, Crossover between Integer and Fractional Vortex Lattices in Coherently Coupled Two-Component Bose-Einstein Condensates, 
Phys. Rev. Lett. {\bf 111}, 170401 (2013). 

\bibitem{nitta2013A} M. Cipriani and M. Nitta, Vortex lattices in three-component Bose-Einstein condensates under rotation: Simulating colorful vortex lattices in a color superconductor, 
Phys. Rev. A {\bf 88}, 013634 (2013). 

\bibitem{dantas} D. S. Dantas, A. R. P. Lima, A. Chaves, C. A. S. Almeida, G. A. Farias, and M. V. Milo{\v s}evi{\'c}, Bound vortex states and exotic lattices in multicomponent Bose-Einstein condensates: The role of vortex-vortex interaction, 
Phys. Rev. A {\bf 91}, 023630 (2015).

\bibitem{kasamatsu2016} K. Kasamatsu, M. Eto, and M. Nitta, Short-range intervortex interaction and interacting dynamics of half-quantized vortices in two-component Bose-Einstein condensates, 
Phys. Rev. A {\bf 93}, 013615 (2016). 

\bibitem{stringari2016} M. Tylutki, L. P. Pitaevskii, A. Recati, and S. Stringari, Confinement and precession of vortex pairs in coherently coupled Bose-Einstein condensates, 
Phys. Rev. A {\bf 93}, 043623 (2016). 

\bibitem{eto2018} M. Eto and M. Nitta, Confinement of half-quantized vortices in coherently coupled Bose-Einstein condensates: Simulating quark confinement in a QCD-like theory, 
Phys. Rev. A {\bf 97}, 023613 (2018). 

\bibitem{lamacraft} B. M. Uranga and A. Lamacraft, Infinite lattices of vortex molecules in Rabi-coupled condensates, 
Phys. Rev. A {\bf 97}, 043609 (2018). 

\bibitem{cross1979} D. L. Stein and M. C. Cross, Phase Transitions in Two-Dimensional Superfluid $^{3}\mathrm{He}$, 
Phys. Rev. Lett. {\bf 42}, 504 (1979). 

\bibitem{korshunov1984} S. E. Korshunov, Two-dimensional superfluid Fermi-liquid with $p$-pairing, 
Zh. Eksp. Teor. Fiz. {\bf 89}, 531 (1985). 

\bibitem{granato1986} E. Granato, J. M. Kosterlitz, and J. Poulter, 
Critical behavior of coupled XY models, 
Phys. Rev. B {\bf 33}, 4767 (1986). 

\bibitem{bighin2019} G. Bighin, N. Defenu, I. N{\'a}ndori, L. Salasnich, and A. Trombettoni, 
Berezinskii-Kosterlitz-Thouless Paired Phase in Coupled XY Models, 
Phys. Rev. Lett. {\bf 123}, 100601 (2019). 

\bibitem{nitta} M. Kobayashi, M. Eto, and M. Nitta, Berezinskii-Kosterlitz-Thouless Transition of Two-Component Bose Mixtures with Intercomponent Josephson Coupling, 
Phys. Rev. Lett. {\bf 123}, 075303 (2019). 

\bibitem{abad2013} M. Abad and A. Recati, A study of coherently coupled two-component Bose-Einstein condensates, 
Eur. Phys. J. D {\bf 67}, 148 (2013).

\bibitem{bertacco2017} A. Cappellaro, T. Macri, G. F. Bertacco, and 
L. Salasnich, Equation of state and self-bound droplet in Rabi-coupled Bose mixtures, 
Sci. Rep. {\bf 7}, 13358 (2017). 

\bibitem{perali2014} A. Guidini and A. Perali, Band-edge BCS-BEC crossover in a two-band superconductor: physical properties and detection parameters, 
Supercond. Sci. Technol. {\bf 27}, 124002 (2014). 

\bibitem{perali2019} L. Salasnich, A. A. Shanenko, A. Vagov, J. Albino Aguiar, and A. Perali, 
Screening of pair fluctuations in superconductors with coupled shallow and deep bands: A route to higher-temperature superconductivity, 
Phys. Rev. B {\bf 100}, 064510 (2019). 

\bibitem{pieri} H. Tajima, Y. Yerin, A. Perali, and P. Pieri, 
Enhanced critical temperature, pairing fluctuation effects, and BCS-BEC crossover in a two-band Fermi gas, 
Phys. Rev. B {\bf 99}, 180503 (2019). 

\bibitem{perali2015} M. V. Milo\v{s}evi{\'c} and A. Perali, 
Emergent phenomena in multicomponent superconductivity: an introduction to the focus issue,
Supercond. Sci. Technol. {\bf 28}, 060201 (2015). 

\bibitem{LeeYang} T. D. Lee and C. N. Yang, Low-Temperature Behavior of a Dilute Bose System of Hard Spheres. II. Nonequilibrium Properties, 
Phys. Rev. {\bf 113}, 1406 (1959). 

\bibitem{griffin} A. Griffin and E. Zaremba, First and second sound in a uniform Bose gas, 
Phys. Rev. A {\bf 56}, 4839 (1997). 

\bibitem{taylor} E. Taylor, H. Hu, X.-J. Liu, L. P. Pitaevskii, A. Griffin, and S. Stringari, 
First and second sound in a strongly interacting Fermi gas, 
Phys. Rev. A {\bf 80}, 053601 (2009). 

\bibitem{hu} H. Hu, E. Taylor, X.-J. Liu, S. Stringari, and A. Griffin, Second sound and the density response function in uniform superfluid atomic gases, 
New J. Phys. {\bf 12}, 043040 (2010). 

\bibitem{stringari} L. Verney, L. Pitaevskii, and S. Stringari, Hybridization of first and second sound in a weakly interacting Bose gas, 
Europhys. Lett. {\bf 111}, 40005 (2015). 

\bibitem{landau} L. D. Landau, Theory of the Superfluidity of Helium II, 
Phys. Rev. {\bf 60}, 356 (1941). 

\bibitem{giamarchi} T. Giamarchi, {\it Quantum Physics in One Dimension}, International Series of Monographs on Physics (Oxford University Press, Oxford, 2003). 

\bibitem{altlandsimons} A. Altland and B. Simons, {\it Condensed Matter Field Theory} (Cambridge University Press, Cambridge, 2010). 

\bibitem{supplement} See Supplemental Material for more details on the statements and the derivations.

\bibitem{nylen} P. Minnhagen and M. Nyl{\'e}n, Charge density of a vortex in the Coulomb-gas analogy of superconducting films, 
Phys. Rev. B {\bf 31}, 5768 (1985). 

\bibitem{stoof} U. Al Khawaja, J. O. Andersen, N. P. Proukakis, and H. T. C. Stoof, Low dimensional Bose gases, 
Phys. Rev. A {\bf 66}, 013615 (2002). 

\bibitem{duan} W. Zhang, G.-D. Lin, and L.-M. Duan, Berezinskii-Kosterlitz-Thouless transition in a trapped quasi-two-dimensional Fermi gas near a Feshbach resonance, 
Phys. Rev. A {\bf 78}, 043617 (2008). 

\bibitem{bighin} G. Bighin and L. Salasnich, Vortices and antivortices in two-dimensional ultracold Fermi gases, 
Sci. Rep. {\bf 7}, 45702 (2017). 

\bibitem{maccari2020} I. Maccari, N. Defenu, L. Benfatto, C. Castellani, and T. Enss, Interplay of spin waves and vortices in the two-dimensional XY model at small vortex-core energy, 
Phys. Rev. B {\bf 102}, 104505 (2020). 

\bibitem{kobayashi2020} M. Kobayashi, G. Fej{\H o}s, C. Chatterjee, and M. Nitta, Vortex confinement transitions in the modified Goldstone model, 
Phys. Rev. Research {\bf 2}, 013081 (2020). 

\bibitem{landau1941} L. D. Landau, Two-fluid model of liquid helium. II, 
J. Phys. USSR {\bf 5}, 71 (USSR) (1941). 

\bibitem{landaufluid} L. D. Landau and E. M. Lifshitz, {\it Fluid Mechanics} (Pergamon, Oxford, 1987). 

\bibitem{khala} I. M. Khalatnikov, {\it An Introduction to the Theory of Superfluidity} (Benjamin, New York, 1965). 

\bibitem{meppelink} R. Meppelink, S. B. Koller, and P. van der Straten, Sound propagation in a Bose-Einstein condensate at finite temperatures, 
Phys. Rev. A {\bf 80}, 043605 (2009). 

\bibitem{tononi} A. Tononi, A. Cappellaro, G. Bighin, and L. Salasnich, Propagation of first and second sound in a two-dimensional Fermi superfluid, 
Phys. Rev. A {\bf 103}, L061303 (2021). 

\bibitem{lellouch2013} S. Lellouch, T.-L. Dao, T. Koffel, and 
L. Sanchez-Palencia, Two-component Bose gases with one-body and two-body couplings, 
Phys. Rev. A {\bf 88}, 063646 (2013). 

\bibitem{search2001} C. P. Search, A. G. Rojo, and P. R. Berman, Ground state and quasiparticle spectrum of a two-component Bose-Einstein condensate, 
Phys. Rev. A {\bf 64}, 013615 (2001). 

\bibitem{Tommasini2003} P. Tommasini, E. J. V. de Passos, 
A. F. R. de Toledo Piza, M. S. Hussein, and E. Timmermans, Bogoliubov theory for mutually coherent condensates, 
Phys. Rev. A {\bf 67}, 023606 (2003). 


\end{thebibliography}

\begin{thebibliography}{999}

\bibitem[S1]{SMpopov} V. N. Popov and A. V. Seredniakov, 
Low-frequency asymptotic  form of the self-energy parts of a  superfluid Bose system at $T=0$, 
Sov. Phys. JETP {\bf 50}, 193 (1979). 

\bibitem[S2]{SMsvistunov} B. Svistunov, E. Babaev, and N. Prokof'ev, {\it Superfluid States of Matter} (CRC Press, Boca Raton, FL, 2015). 

\bibitem[S3]{SMgiamarchi} T. Giamarchi, {\it Quantum Physics in One Dimension}, International Series of Monographs on Physics (Oxford University Press, Oxford, 2003). 

\bibitem[S4]{SMaltlandsimons} A. Altland and B. Simons, {\it Condensed Matter Field Theory} (Cambridge University Press, Cambridge, 2010). 

\bibitem[S5]{SMnelson} D. R. Nelson and J. M. Kosterlitz, Universal Jump in the Superfluid Density of Two-Dimensional Superfluids, 
Phys. Rev. Lett. {\bf 39}, 1201 (1977). 

\bibitem[S6]{SMkosterlitz} J. M. Kosterlitz and D. J. Thouless, Ordering, metastability and phase transitions in two-dimensional systems, 
J. Phys. C {\bf 6}, 1181 (1973). 

\bibitem[S7]{SMenss} V. Karle, N. Defenu, and T. Enss, Coupled superfluidity of binary Bose mixtures in two dimensions, 
Phys. Rev. A {\bf 99}, 063627 (2019). 

\bibitem[S8]{SMson2002} D. T. Son and M. A. Stephanov, Domain walls of relative phase in two-component Bose-Einstein condensates, 
Phys. Rev. A. {\bf 65}, 063621 (2002). 

\bibitem[S9]{SMfurutani} K. Furutani, A. Tononi, and L. Salasnich, Sound modes in collisional superfluid Bose gases, 
New. J. Phys. {\bf 23}, 043043 (2021). 

\bibitem[S10]{SMlandau1941} L. D. Landau, Two-fluid model of liquid helium. II, 
J. Phys. USSR {\bf 5}, 71 (1941). 

\end{thebibliography}
\end{document}